\documentclass[sigplan,screen]{acmart}

\copyrightyear{2026}
\acmYear{2026}
\setcopyright{cc}
\setcctype{by}
\acmConference[ASPLOS '26]{Proceedings of the 31st ACM International Conference on Architectural Support for Programming Languages and Operating Systems, Volume 2}{March 22--26, 2026}{Pittsburgh, PA, USA}
\acmBooktitle{Proceedings of the 31st ACM International Conference on Architectural Support for Programming Languages and Operating Systems, Volume 2 (ASPLOS '26), March 22--26, 2026, Pittsburgh, PA, USA}
\acmPrice{}
\acmDOI{10.1145/3779212.3790197}
\acmISBN{979-8-4007-2359-9/2026/03}



\usepackage[]{hyperref}
\usepackage{colortbl}
\usepackage{makecell}
\usepackage{footnote}
\usepackage{multirow}
\usepackage{enumitem}

\begin{document}

\title[Ouroboros]{Ouroboros: Wafer-Scale SRAM CIM with Token-Grained Pipelining for Large Language Model Inference}

\author{Yiqi Liu}
\orcid{0009-0005-8717-068X}
\affiliation{
  \institution{SKLP, Institute of Computing Technology, Chinese Academy of Sciences}
  \city{Beijing}
  \country{China}
  \\
  \institution{University of Chinese Academy of Sciences}
  \city{Beijing}
  \country{China}
}
\email{liuyiqi23@mails.ucas.ac.cn}
\authornote{Equal contribution.}

\author{Yudong Pan}
\orcid{0009-0001-0012-4113}
\affiliation{
  \institution{SKLP, Institute of Computing Technology, Chinese Academy of Sciences}
  \city{Beijing}
  \country{China}
  \\
  \institution{University of Chinese Academy of Sciences}
  \city{Beijing}
  \country{China}
}
\email{panyudong23@mails.ucas.ac.cn}
\authornotemark[1]

\author{Mengdi Wang}
\orcid{0000-0002-7012-2308}
\affiliation{
  \institution{SKLP, Institute of Computing Technology, Chinese Academy of Sciences}
  \city{Beijing}
  \country{China}
}
\email{wangmengdi@ict.ac.cn}
\authornote{Corresponding authors.}

\author{Shixin Zhao}
\orcid{0000-0002-5175-7025}
\affiliation{
  \institution{SKLP, Institute of Computing Technology, Chinese Academy of Sciences}
  \city{Beijing}
  \country{China}
  \\
  \institution{University of Chinese Academy of Sciences}
  \city{Beijing}
  \country{China}
}
\email{zhaoshixin18@mails.ucas.ac.cn}

\author{Haonan Zhu}
\orcid{0000-0002-5644-3105}
\affiliation{
  \institution{SKLP, Institute of Computing Technology, Chinese Academy of Sciences}
  \city{Beijing}
  \country{China}
  \\
  \institution{Hangzhou Institute for Advanced Study, University of Chinese Academy of Sciences}
  \city{Hangzhou}
  \country{China}
}
\email{zhuhaonan24@mails.ucas.ac.cn}

\author{Yinhe Han}
\orcid{0000-0003-0904-6681}
\affiliation{
  \institution{SKLP, Institute of Computing Technology, Chinese Academy of Sciences}
  \city{Beijing}
  \country{China}
}
\email{yinhes@ict.ac.cn}

\author{Lei Zhang}
\orcid{0000-0001-9711-8758}
\affiliation{
  \institution{SKLP, Institute of Computing Technology, Chinese Academy of Sciences}
  \city{Beijing}
  \country{China}
}
\email{zlei@ict.ac.cn}

\author{Ying Wang}
\orcid{0000-0001-5172-4736}
\affiliation{
  \institution{SKLP, Institute of Computing Technology, Chinese Academy of Sciences}
  \city{Beijing}
  \country{China}
}
\email{wangying2009@ict.ac.cn}
\authornotemark[2]

\renewcommand{\shortauthors}{Yiqi Liu et al.}

\begin{abstract}

Large language model (LLM) inference demands vast memory capacity and hierarchical memory structures, but conventional architectures suffer from excessive energy and latency costs due to frequent data movement across deep memory tiers. To address this, we propose a wafer-scale SRAM-based Computing-in-Memory (CIM) architecture that performs all LLM operations in situ within the first-level SRAM, eliminating off-chip data migration and achieving unprecedented energy efficiency. However, wafer-scale SRAM CIM presents multiple challenges due to the limited first-level memory capacity, which requires efficient compute-memory resource allocation.

To enable efficient LLM execution on this architecture, we propose three key innovations: 
\textbf{Token-Grained Pipelining} – Conventional sequence-level pipelining suffers from underutilization due to varying input sequence lengths and batching policies. We introduce a fine-grained token-level pipeline that mitigates sequence length variations, enhancing the utilization of CIM cores while minimizing the storage capacity required for activation.
\textbf{Distributed Dynamic KV Cache Management} – KV cache storage occupies significant memory in LLM inference. We optimize on-chip KV caching by decoupling CIM memory from compute assignment, leveraging fragmented SRAM CIM memory within already-allocated cores for efficient KV storage, and reducing dedicated memory overhead.
\textbf{Communication-Aware and Fault-Tolerant Core Mapping} – Efficient execution on wafer-scale CIM requires optimal mapping from transformer blocks to CIM cores to minimize inter-(pipeline) stage communication while ensuring intra-stage compute locality. We design a network-on-wafer-aware mapping strategy that places pipeline stages in close proximity while also distributing large layers efficiently across multiple cores. This mapping accounts for core-level defects, improving fault tolerance in wafer-scale deployment.

Experimental results demonstrate that Ouroboros achieves $4.1\times$ average throughput improvement and $4.2\times$ average energy efficiency gain over state-of-the-art systems, peaking at $9.1\times$ throughput and $17\times$ energy efficiency for the 13B model.
\end{abstract}

\maketitle 

\section{Introduction}

Transformer-based Large Language Models (LLMs)~\cite{avaswani2017attention} have demonstrated unprecedented performance across a wide range of natural language processing tasks. This success is largely attributed to their increasing scale, with models now reaching hundreds of billions of parameters~\cite{touvron2023llama,brown2020language,chowdhery2023palm}. However, the rapid growth in LLM size has led to a fundamental deployment challenge: LLM inference is becoming increasingly constrained by \textbf{memory capacity, memory bandwidth, and data movement overheads.} As LLMs scale, a significant fraction of inference energy is wasted on shuffling model parameters, activations, and Key-Value (KV) data across different levels of the memory hierarchy ~\cite{dao2022flashattention}.

\begin{figure}[htbp]
    \begin{center}
    \includegraphics[width=0.47\textwidth]{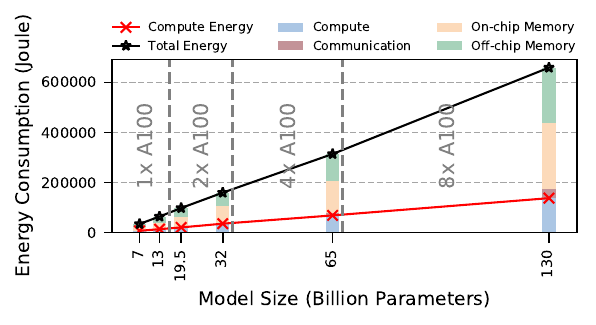}
    \end{center}
    \caption{Hardware Scaling Tax Due to Increasing Model Size.}
    \label{fig:Tax}
\end{figure}

Current LLM inference architectures rely on deep memory hierarchies (e.g., first-level on-chip SRAM and second-level off-chip DRAM/High Bandwidth Memory (HBM)~\cite{nvidia_a100,nvidia_h100}). However, as illustrated in Fig.~\ref{fig:Tax}, this technical route incurs \textbf{"hardware scaling tax"}—wherein data moving in between deep memory hierarchy and between-memory interconnect traversals dominate energy consumption, far outweighing the cost of computation. Given that LLM inference is memory-intensive, this excessive data movement fundamentally limits energy efficiency and throughput.

To address this challenge, we propose Ouroboros, a \textbf{wafer-scale SRAM-based Computing-in-Memory (CIM)} architecture, for LLM inference. This architecture performs all LLM operations in the first-level SRAM arrays to eliminate data migration across the deep memory hierarchy. 
Due to the exclusive adoption of low-density SRAM, Ouroboros faces inherent capacity constraints that necessitate high SRAM utilization. 
In CIM architecture, SRAM inherently serves dual functions of both storage and computation, necessitating the optimization of both storage and computation utilization. This introduces multiple key challenges:

\textbf{Challenge \#1: Low Utilization Due to Pipeline Bubble.}
Pipeline parallelism enhances activation storage efficiency by immediately consuming intermediate activation. However, this approach suffers from \textbf{suboptimal CIM utilization} because the computational intensity of LLM inference is highly variable—fluctuating with input sequence lengths and batching policies. 
The conventional sequence-grained pipeline, where each stage processes different sequences with varying lengths, introduces load imbalance across stages~\cite{park2024attacc, aminabadi2022deepspeed}. This results in
\textbf{pipeline bubbles}, where some CIM cores remain idle while others are overloaded, leading to inefficient memory and compute resource utilization. To address this, we propose a \textbf{token-grained pipelining (TGP) architecture} that partitions sequences into token-level units and schedules them across different pipeline stages, mitigating the dynamic variations in computational intensity caused by variable input lengths and distinct inference phases.
This approach reduces the memory footprint of intermediate activations and prevents CIM core underutilization, thereby maximizing throughput and memory efficiency.

\textbf{Challenge \#2: Low Utilization Due to KV Storage.}
KV cache storage is also a major bottleneck in LLM inference, requiring large memory spaces and dynamic management. How to utilize the limited on-chip SRAM space to accommodate KV caching is a significant challenge. 
Traditional centralized dynamic management methods are not suitable for large-scale distributed CIM core arrays and fail to address the load balancing of CIM cores~\cite{kwon2023efficient, 298683}.
We propose a distributed dynamic KV cache management strategy that repurposes underutilized memory in CIM cores already allocated to transformer blocks. Furthermore, we introduce a corresponding KV mapping method to ensure load balancing. Together, these approaches significantly enhance the utilization efficiency of CIM resources.

\textbf{Challenge \#3: Low Utilization Due to Communication Overhead.}
In CIM architectures, computational operations are inherently restricted to the physical locations where weights and KV data are stored, making data communication a bottleneck that limits computational utilization.
In a wafer-scale chip (WSC) with thousands of CIM cores distributed across a massive die, this in-place computation paradigm creates a fundamental trade-off between inter-layer communication (between pipeline stages) and intra-layer communication (within a stage for reductions). 
Specifically, if cores belonging to the same pipeline stage are tightly collocated, far inter-stage communication incurs significant overhead. Conversely, if cores belonging to different pipeline stages are interleaved across the wafer, intra-stage reduction introduces substantial overhead.
Striking a balance between minimizing inter-layer traversal and ensuring intra-layer compute locality is critical for optimizing performance.

To resolve this, we employ Mixed Integer Quadratic Programming (MIQP) to efficiently map transformer blocks onto WSC, ensuring that neighboring pipeline stages are placed optimally to minimize intra-layer and inter-layer communication. Additionally, we leverage Dynamic Programming (DP) to mitigate intra-core communication pressure, ensuring that workloads are distributed efficiently within each CIM core. 
Moreover, our mapping strategy is fault-tolerant, accounting for manufacturing defects in the wafer-scale CIM fabric and intelligently redistributing workloads to ensure robust execution and reliability.

To summarize, the key contributions of this paper are as follows:
\begin{itemize}[leftmargin=*]
\item We introduce Ouroboros, a novel wafer-scale SRAM-based CIM architecture to minimize data movement, thereby achieving low energy consumption and high throughput.
\item We propose a novel TGP strategy that enhances CIM core utilization by reducing bubbles.
\item We introduce a distributed dynamic KV cache management scheme, along with a corresponding KV distribution strategy, to enhance the utilization of CIM resources.
\item We design an optimized CIM core mapping strategy that minimizes inter-core communication latency to enhance the utilization of computational resources while ensuring fault tolerance for wafer-scale LLM inference.
\item We conducted experiments across multiple models and performed multi-wafer scalability tests. Compared to state-of-the-art (SOTA) architectures, we achieved an average performance improvement of $4.1\times$ and an energy efficiency enhancement of $4.2\times$.
\end{itemize}

\begin{figure*}[htbp]
    \begin{center}
    \includegraphics[width=0.87\textwidth]{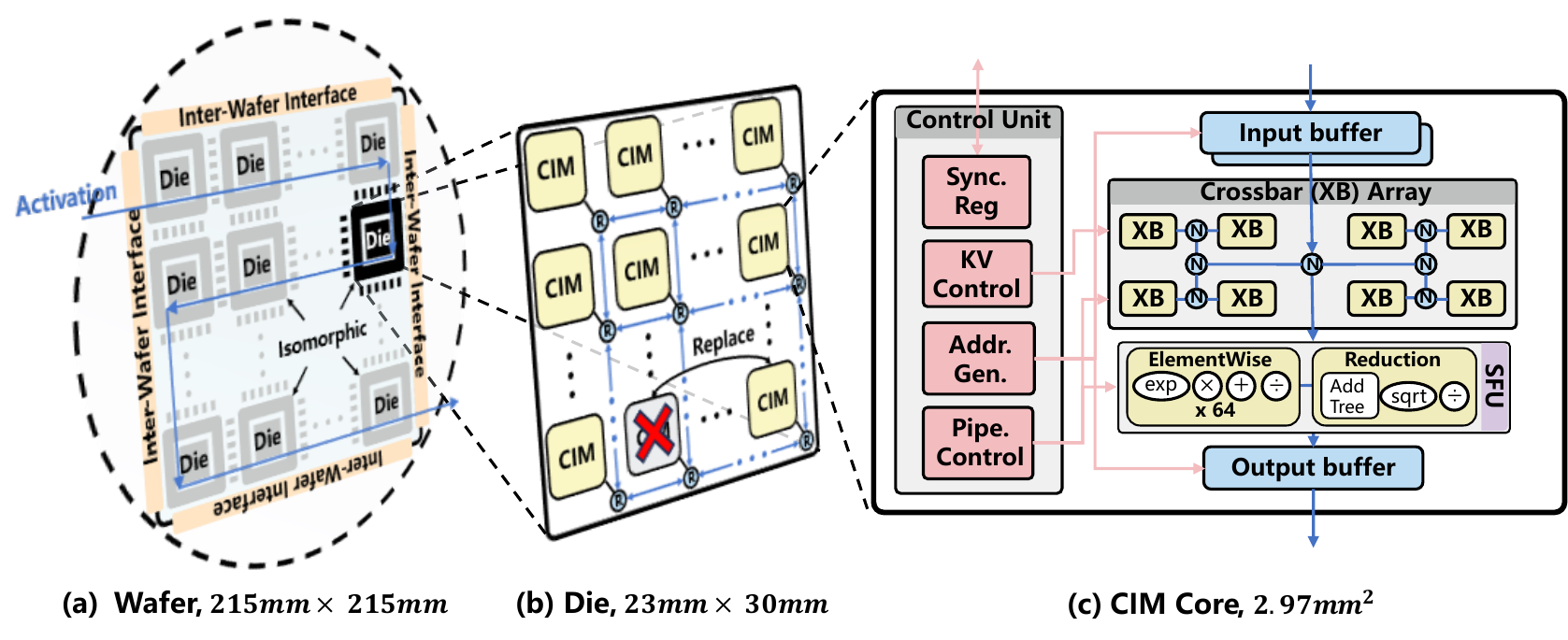}
    \end{center}
    \caption{Hierarchical Architecture of Ouroboros.}
    \label{fig:overall_architecture}
\end{figure*}

\section{Background}

\subsection{Large Language Model Inference}
LLMs comprise N identical transformer blocks, each integrating a multi-head attention (MHA) mechanism and a feed-forward network (FFN). The MHA projects input sequences into Query/Key/Value (Q/K/V) matrices per attention head, computes attention scores via \( S = QK^T \), applies softmax, and outputs \( \text{softmax}(S)V \). These head outputs are subsequently concatenated and projected. The FFN follows with two fully-connected layers: FFN1 upscales activation dimensions while FFN2 restores the original dimensionality. 

LLM inference consists of two phases: \textbf{prefill} and \textbf{decoding}. In the prefill phase, the entire input sequence 
(e.g., a user prompt)
is processed in parallel to produce hidden representations and cache KV pairs for later decoding. This phase is compute-intensive but occurs only once per request. The decoding phase then generates output tokens autoregressively using the KV cache: at each step, attention is computed only for the newest token, and the cache is updated. This iterative process is memory-bound, with memory access overhead growing linearly with the output sequence length.

\textbf{Computational and Memory Challenges.}
As LLMs grow in size, inference efficiency is increasingly dominated by memory bandwidth and data movement overheads rather than raw computation. Deep memory hierarchies (e.g., HBM and DRAM) introduce significant energy costs due to frequent weight and KV cache accesses, while multi-hop interconnects in distributed systems further exacerbate communication latency. These inefficiencies motivate a novel architecture that eliminates memory hierarchies and performs all LLM operations in situ within high-speed SRAM.

\subsection{Wafer-Scale Chip}
WSCs leverage advanced packaging for scaling up chip size, delivering superior computational power and efficiency. By eliminating chip-to-chip interconnects, they achieve significantly lower latency, higher bandwidth, and reduced power consumption versus multi-die alternatives—making them ideal for massively parallel applications including artificial intelligence, high-performance computing, and large-scale data analytics ~\cite{jiang2020cu,james2020ispd,chun2020info_sow}.
However, efficiently mapping LLM execution onto a WSC requires overcoming key architectural challenges, such as scaling SRAM-based storage, optimizing pipelining for high core utilization, dynamic allocation of KV caches within distributed storage systems, and minimizing inter-core communication overhead.

\subsection{Computing-in-Memory}
SRAM-based CIM architectures are widely used for AI acceleration due to their high-speed access, low-latency compute capabilities, and energy efficiency~\cite{asifuzzaman2023survey}. By merging memory and logic functions, they enable in situ operations that significantly reduce data movement between memory and processors~\cite{gomez2022benchmarking,jhang2021challenges,kwon202125,su2021two}. 
However, SRAM suffers from low density compared to DRAM, meaning that traditional SRAM-based CIM cannot accommodate the multi-gigabyte-scale storage demands of LLMs. Without sufficient on-chip memory, frequent off-chip DRAM accesses negate CIM's energy benefits.

\textbf{Wafer-Scale CIM for LLM Inference.}
To avoid deep-tier memory access, we propose a wafer-scale SRAM CIM architecture that distributes LLM model parameters, KV cache, and activations across thousands of CIM cores. This unified SRAM CIM approach eliminates memory hierarchy-induced data movement, achieving maximum energy efficiency. However, making this architecture viable requires addressing the following:

\begin{itemize}[leftmargin=*]
\item Efficient SRAM utilization via TGP to prevent CIM core underutilization.
\item Optimized KV cache management to allocate fragmented memory across CIM cores dynamically.
\item Hierarchical core mapping strategies to balance intra-layer and inter-layer communication while ensuring fault tolerance.
\end{itemize}

\section{Architecture Overview}
\label{section:Architecture_Overview}

The Ouroboros wafer-scale SRAM CIM system is structured hierarchically into three levels:

\textbf{Wafer-Scale Integration.}
The top-level structure is a monolithic WSC measuring 215mm × 215mm (Fig.~\ref{fig:overall_architecture}(a)), fabricated on a 12-inch wafer. The system integrates 54GB of on-chip SRAM to store weights, activation, and KV entirely in the first-level memory, eliminating costly DRAM accesses.

The wafer is composed of 9 rows × 7 columns of homogeneous dies, each measuring 23mm × 30mm, forming a seamless interconnected compute fabric. They are directly interconnected via stitching technology using offset exposures and form a mesh topology.
For scalable inter-die dataflow, an S-shaped logical routing topology adapts to the pipelines' producer-consumer model, enabling sequential activation propagation while optimizing inter-die bandwidth.
To support models that exceed the capacity of a single wafer, we employ the optical Ethernet port to achieve multi-wafer scaling.

\textbf{Die-Level Organization.}
\label{section:Die_Level_Organization}
Each die comprises a 13 × 17 grid of CIM cores interconnected in a high-bandwidth mesh network (Fig. \ref{fig:overall_architecture}(b)). The die area is pushed close to the reticle limit to maximize on-die SRAM capacity while minimizing inter-die communication overhead. To ensure reliability despite large die sizes, we incorporate core-level fault tolerance, dynamically remapping workloads to bypass defective cores (detailed in Section \ref{section:Fault}).

Each core-to-core link provides 256-bit bidirectional bandwidth—matching the core’s buffer width—in all four directions, enabling high-throughput intra-die data transfers critical for both intra-layer reductions and inter-layer communication.

\textbf{CIM Core Microarchitecture.}
The structure of a single CIM core is shown in Fig. \ref{fig:overall_architecture} (c). The $2.97mm^2$ Core contains the following components: 1) Input (128KB for ping-pong) and output (32KB) buffers for storing 8-bit intermediate activation. This capacity accommodates full token (tens of KB) for contemporary LLMs~\cite{touvron2023llama,yang2024qwen2} and long-sequence scores, effectively mitigating inter-core 32-bit partial sum transfers caused by input channel partitioning. Each buffer delivers 256-bit/cycle bandwidth, matching the crossbar's native parallelism. 2) A 32-crossbar array (detailed implementation in Section \ref{section:CIM_Implementation}) with 4MB SRAM capacity balances activation broadcasting requirements for small cores against pipeline imbalance and resource underutilization in large cores. These arrays interconnect via a 1024-bit H-tree-based NoC. Quadruple the link width is provisioned to accommodate the accumulation of partial sums with 4$\times$ input width. 3) A special function unit (SFU) executes softmax and similar operations, integrating 64-way parallel ElementWise and Reduction units with a 10KB buffer. This parallelism enables pipeline throughput matching between the SFU and crossbar array. 4) A control unit coordinates synchronization across multiple CIM cores for both model parallelism and pipeline parallelism, while orchestrating the pipeline between internal crossbars and SFU.

\section{End-to-End Inference Framework}

\subsection{Framework Overview}

\begin{figure}[htbp]
    \begin{center}
    \includegraphics[width=0.38\textwidth]{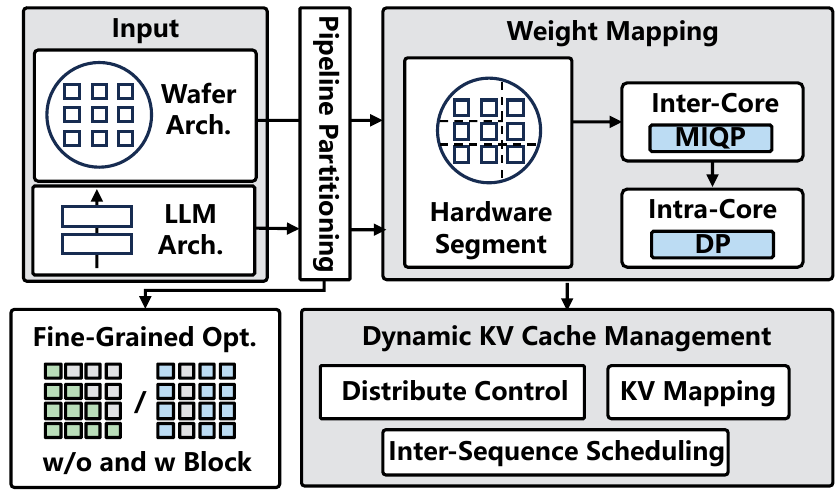}
    \end{center}
    \caption{Overview of End-to-End Inference Framework.}
    \label{fig:ruan_arch}
\end{figure}

We proposed an end-to-end (E2E) inference framework for LLMs, as detailed in Fig. \ref{fig:ruan_arch}. This framework begins with the pipeline partitioning of the model and proceeds with fine-grained pipeline optimizations aimed at optimizing CIM resource utilization. Subsequently, we devised a weight mapping strategy. This encompasses hardware resource allocation and a hierarchical mapping methodology for minimizing communication. Finally, to enhance the utilization of the on-chip KV cache, we developed a distributed dynamic KV cache management scheme and corresponding hardware support.

\subsection{Pipeline Parallelism for High CIM Utilization}
\label{section:Scheduling}

\begin{figure}[htbp]
    \begin{center}
    \includegraphics[width=0.38\textwidth]{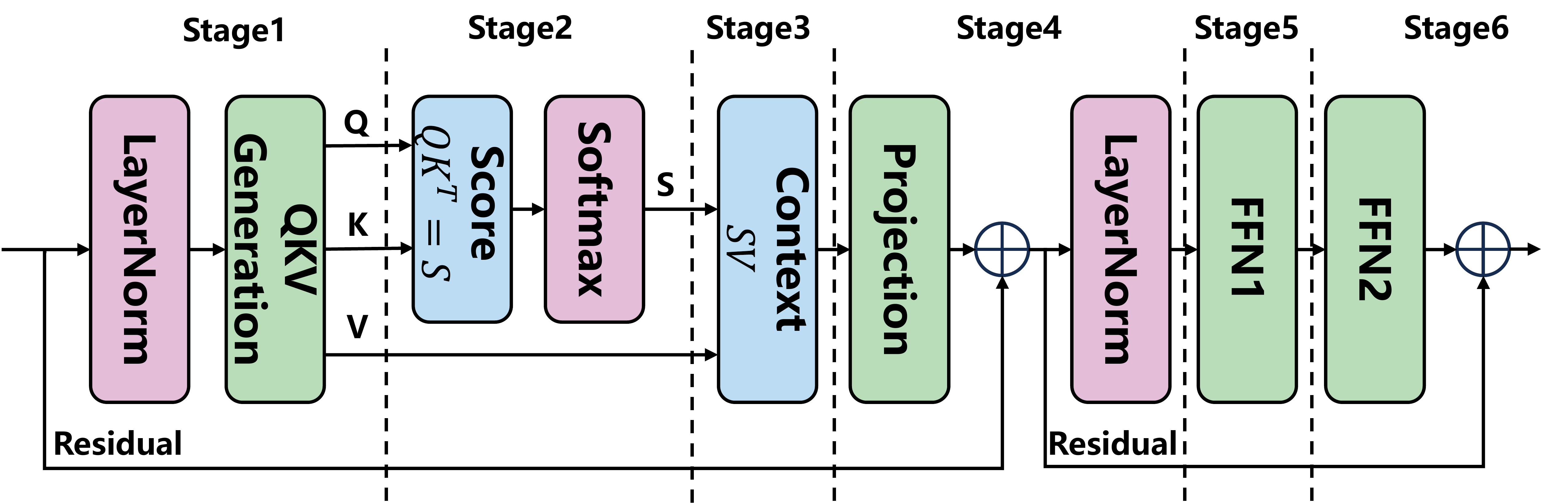}
    \end{center}
    \caption{Pipeline Stage Partitioning of Transformer Blocks.}
    \label{fig:pipe_stage}
\end{figure}

To fully utilize all CIM cores in parallel, we adopted a fully unrolled pipeline enabling concurrent execution of operators with inter-dependencies. As shown in Fig. \ref{fig:pipe_stage}, each transformer block is split into six stages, resulting in a unified 6N-stage pipeline for a model with N transformer blocks.

\subsubsection{Token-Grained Pipeline}
\label{section:TGP}
\begin{figure}[htbp]
    \begin{center}
    \includegraphics[width=0.40\textwidth]{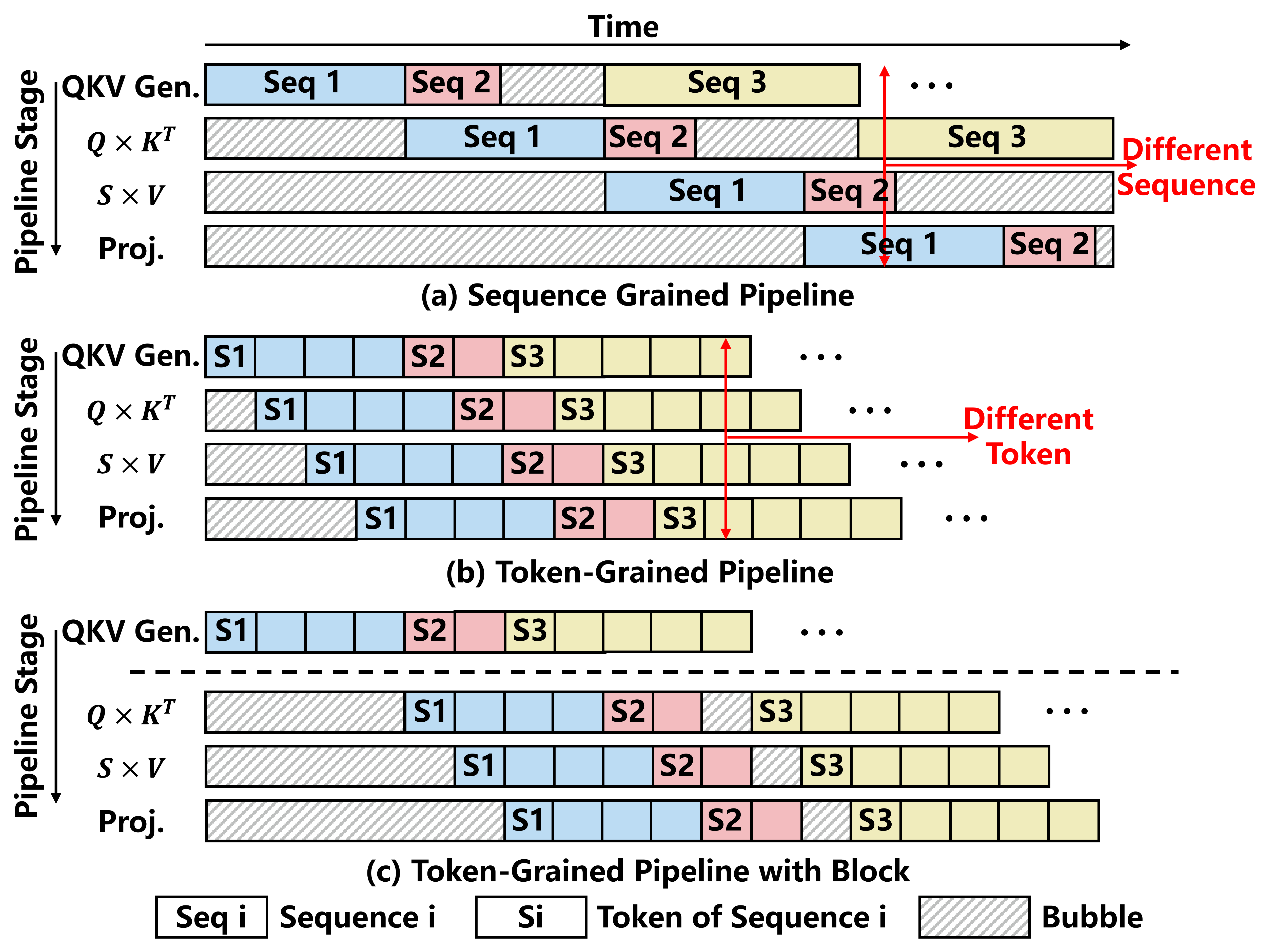}
    \end{center}
    \caption{Spatial-Temporal Diagram of Different Pipeline Strategies.}
    \label{fig:TGP}
\end{figure}

Current pipelining strategies, whether at the mini-batch level or the head/attention level, operate with sequences as the minimum granularity~\cite{park2024attacc, aminabadi2022deepspeed}, where different sequences are processed at distinct pipeline stages, as shown in Fig. \ref{fig:TGP} (a). 
The inference service dynamically receives requests of varying lengths, and this dynamism leads to imbalances in statically allocated pipelines, inducing severe pipeline bubbles that significantly degrade CIM resource utilization. In a system like Ouroboros, which is unable to disaggregate the prefill and decoding phases~\cite{zhong2024distserve} due to memory constraints, this imbalance is further exacerbated when the batching policy~\cite{yu2022orca} schedules them together.

To solve this problem, we propose the novel TGP. Unlike sequence-level pipelining, TGP employs a single token as the smallest processing unit to avoid the imbalance of varying sequence lengths. 
Such fine-grained partitioning eliminates weight reuse, incurring significant energy overhead from repeated weight shuffling across memory hierarchies. WSC enables full on-chip pipelined execution of LLMs to avoid DRAM access, while synergistic CIM integration further eliminates SRAM data movement.
However, attention mechanisms require simultaneous QKV computation for all tokens. Since TGP generates QKVs sequentially, this creates pipeline stalls until full sequence generation completes.

Considering that most LLMs are decoder-only architectures, they apply a causal mask to the S matrix to enforce that each input token performs attention calculations only with itself and preceding tokens, as shown in Fig. \ref{fig:qiguai_mask} (a). 
Hence, the effective values of the S matrix are distributed in the lower triangular region, with the attention computations of the current token for subsequent tokens being discarded. Therefore, in the prefill phase, as each token's QKV values are generated, attention calculations can be performed between the current token and all preceding tokens without waiting for the KV pairs of all tokens to be fully generated. This approach aligns with the token-level pipeline granularity and ensures equivalent computational results. In the decoding phase, this issue is mitigated because autoregressive generation processes one token at a time.

Through this fine-grained pipeline, as shown in Fig. \ref{fig:TGP} (b), different pipeline stages operate on different tokens rather than different sequences. 
In this scenario, the computational load remains uniformly balanced across all pipeline stages, as each stage processes a single token with no dynamic variation in computational intensity, effectively mitigating the issue of low resource utilization caused by pipeline stalling.

Besides, TGP reduces storage capacity requirements. By refining the pipeline granularity to a single token, the intermediate activation buffer stores only individual tokens instead of entire sequences. For the context window size of common LLMs today, the storage of activation is reduced by a factor of thousands~\cite{touvron2023llama,brown2020language,chowdhery2023palm}.

\subsubsection{Adaptation of Encoder}
\label{section:HGP}

\begin{figure}[htbp]
    \begin{center}
    \includegraphics[width=0.36\textwidth]{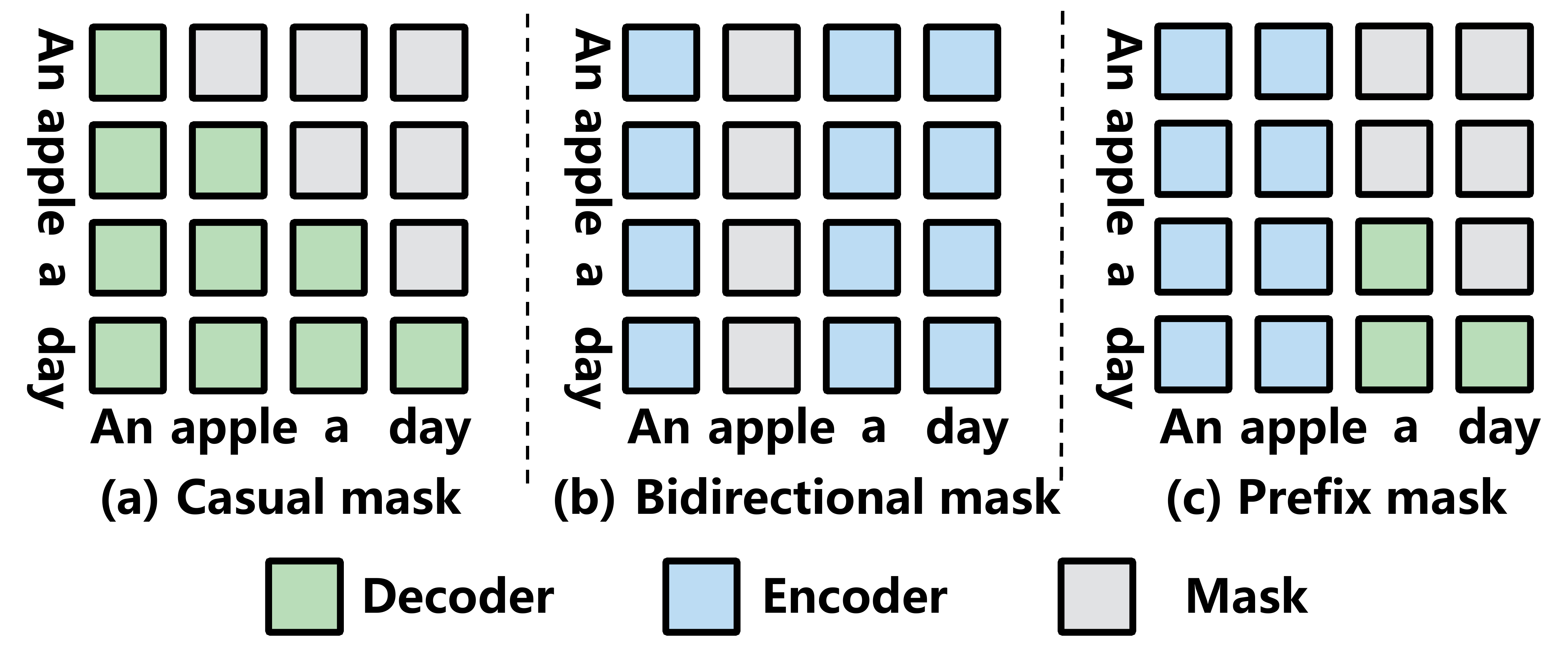}
    \end{center}
    \caption{Forms of Masking in Different LLM Architectures.}
    \label{fig:qiguai_mask}
\end{figure}

For LLMs with other architectures, such as bidirectional attention and seq2seq~\cite{devlin2018bert, raffel2020exploring}, the straightforward application of TGP is not feasible. As illustrated in Fig. \ref{fig:qiguai_mask} (b)(c), the shape of the mask is no longer a perfect inverted triangle. Hence, for each token, attention calculations need to be performed not only with preceding tokens but also with subsequent tokens. This necessitates pipeline stalling until subsequent tokens complete computation.  As shown in Fig. \ref{fig:TGP} (c), attention stages degrade to sequence-level granularity while other stages maintain token-level pipelining.
This incurs minimal throughput degradation since bubbles occur exclusively at sequence-level partitioning boundaries. Only when a newly scheduled sequence exceeds the length of the prior longest sequences does partitioning introduce bubbles equivalent to their length differential.

\begin{figure}[htbp]
    \begin{center}
    \includegraphics[width=0.44\textwidth]{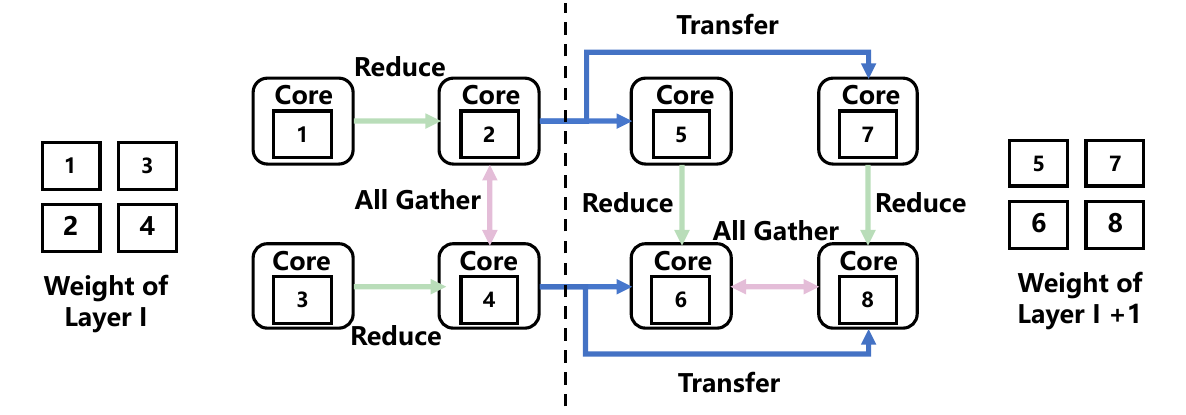}
    \end{center}
    \caption{Inter-Layer Communication and Intra-Layer Communication.}
    \label{fig:tiling}
\end{figure}

\subsection{Communication-Aware Weight Mapping}
\label{section:Mapping}

As shown in Fig. \ref{fig:tiling}, tiling neural network layers across distributed compute cores necessitates inter-core communication for both inter-layer intermediate activations and intra-layer reductions. Co-locating cores processing the same layer minimizes intra-layer communication overhead but requires long-distance inter-layer data transfers, whereas interleaving cores across layers reduces inter-layer routing distances but exacerbates intra-layer communication overhead. Therefore, it is necessary to devise a comprehensive mapping methodology to resolve this conflict.

\subsubsection{Inter-Core Mapping}
\label{section:inter_map}
The inter-core mapping algorithm allocates CIM cores for each layer. Leveraging the properties of LLM and the characteristics of the hardware architecture, we impose the following constraints on the mapping space to reduce its complexity:
(1) LLM typically consists of multiple identical transformer blocks. We only need to perform the mapping of a single transformer block to a group of CIM
 core, and then repeat this mapping. 
(2) For weight tiling, we consistently prioritize partitioning along the output channel dimension to minimize high-bitwidth partial sum transfers. 
We model the search of the mapping space as an MIQP problem~\cite{wolsey2007mixed, achterberg2013mixed}. 

\textbf{Symbolic.}
The symbols used in this problem are as shown in Table \ref{table:sb}. The constant $output(l)$, $reduction(l)$, $gather(l)$, and $I(l),O(l)$ are determined by network attribute and constraint (2). The constant $\#Core(l)$ is derived from the size of the model weights for that layer divided by the storage capacity of the CIM core. Finally, the parameter $Cost_{inter}$ is derived from the intra-die bandwidth divided by the inter-die bandwidth.

\textbf{Objective.}
The primary objective in inter-core mapping is to minimize data transmission between cores: 

\begin{equation}
    \begin{gathered}
        \label{equ:obj}        
        min\ \vec{M}^TQ\vec{M} \\ 
        Q[l_1,i_1,o_1,n_1][l_2,i_2,o_2,n_2]= \\     
        \left\{
        \begin{array}{l}      
        Manh(n_1,n_2) \times output(l) \times Penalty(n_1,n_2), \\
        \qquad if\ \ l_1 + 1 = l_2\ \&\&\ o_1 = i_2\\       
        Manh(n_1,n_2) \times reduce(l) \times Penalty(n_1,n_2), \\
        \qquad if\ \ l_1 = l_2\ \&\&\ o_1 = o_2\ \&\&\ i_2 = I(l_2)\\     
        Manh(n_1,n_2) \times gather(l) \times Penalty(n_1,n_2)\\
        \qquad if\ \ l_1 = l_2\ \&\&\ i_1 = I(l_1)\ \&\&\ i_2 = I(l_2)\\      
        0,\qquad otherwise
        \end{array} \right.       
    \end{gathered}
\end{equation}

In the formula, Manh() represents the Manhattan distance between two cores, $n_1$ and $n_2$. Penalty() denotes the penalty coefficient for cross-die transmission. If $n_1$ and $n_2$ belong to two different dies, then this value is $Cost_{inter}$. Otherwise, it is 1. 

\textbf{Constraints.}
During the mapping process, there are several conditions that must be adhered to, which we formalize as constraints in the MIQP problem. Each core can process at most one tile, and defective cores cannot handle tasks (Eq. \ref{equ:cons1}). The number of cores processing each layer must be the same as pre-determined (Eq. \ref{equ:cons2}).

\begin{equation}
    \begin{gathered}
        \label{equ:cons1}
        \forall{n}\sum_{l=1}^{L}{\sum_{o=1}^{O(l)}{\sum_{i=1}^{I(l)}{M_{l,i,o,n}}}} \leq 1 - D(n)
    \end{gathered}
\end{equation}

\begin{equation}
    \begin{gathered}
        \label{equ:cons2}
        \forall{l}\sum_{n=1}^{N}{\sum_{o=1}^{O(l)}{\sum_{i=1}^{I(l)}{M_{l,i,o,n}}}} = \#Core(l)
    \end{gathered}
\end{equation}

\begin{table}[htbp]
	\caption{Symbolic Table}
	\begin{center}
		\begin{tabular}{ c | p{5.3cm} }
			\hline
                \rowcolor{gray!20}
			\multicolumn{2}{c}{\textbf{Indices}} \\
			\hline
                $l$ & The $l$-th layer in the Transformer block.\\
                \hline
                $i, o$ & The $i/o$-th part of the input/output channels of the tiled layer.\\
                \hline
                $n$ & The $n$-th core awaiting allocation.\\
                \hline
                \rowcolor{gray!20}            \multicolumn{2}{c}{\textbf{Variables}} \\
                \hline
                $M_{l,i,o,n}$ & If the value is 1, it indicates that the core-$n$ is responsible for processing the $i/o$-th part of the input/output channels of the tiled layer $l$; otherwise, it is 0. \\
                \hline
                \rowcolor{gray!20}            \multicolumn{2}{c}{\textbf{Constants}} \\
                \hline
                $output(l)$ & The volume of output activation in the $l$-th layer. \\
                \hline
                $reduction(l)$ & The reduction volume of  the $l$-th layer. \\
                \hline
                $gather(l)$ & The gather volume of the $l$-th layer's. \\
                \hline
                $I(l),O(l)$ & The number of splits along the input and output channels of the $l$-th layer. \\
                \hline
                $\#Core(l)$ & The quantity of cores engaged in the computation of the $l$-th layer. \\
                \hline
                $Cost_{inter}$ & The penalty for die-to-die (D2D) transmission. \\
                \hline
                $D(n)$ & If the value is 1, it indicates that the core-$n$ is defective; otherwise, it is 0. \\
                \hline
		\end{tabular}
		\label{table:sb}
	\end{center}
\end{table}

\subsubsection{Intra-Core Mapping}
\begin{figure}[htbp]
    \begin{center}
    \includegraphics[width=0.38\textwidth]{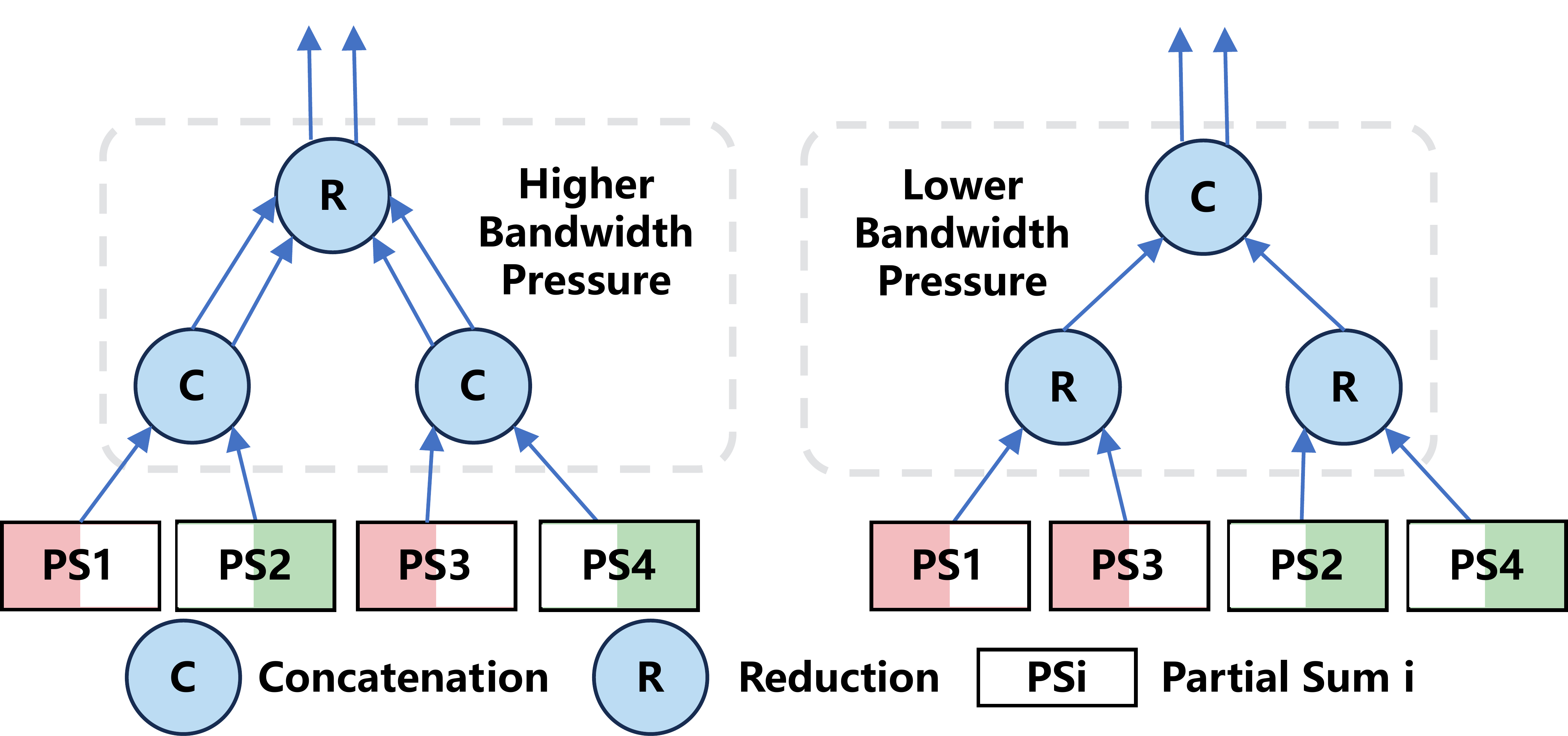}
    \end{center}
    \caption{Binary Tree Abstraction in Intra-Layer Mapping.}
    \label{fig:Tree_Aloc}
\end{figure}

We employ H-Tree topology with 1024-bit width at each level, since reduction halves the data volume before propagating upwards, maintaining consistent bandwidth requirements per level. Based on this interconnection structure, the crossbar array can be abstracted as a binary tree, where the leaf nodes are crossbars and the non-leaf nodes are convergence nodes of the H-Tree. As shown in Fig. \ref{fig:Tree_Aloc}, the problem thus transforms into how to allocate weights to the leaf nodes of the binary tree. The allocation results will determine whether each non-leaf node performs a reduction or concatenation. Performing concatenation operations near the leaf nodes imposes significant communication overhead on these tree segments, as concatenation doubles the dimension of the partial sum.
Therefore, weight distribution aims to ensure that operations near the leaf nodes are predominantly reduction, while concatenation is concentrated closer to the root of the tree. To address this issue, we design a DP algorithm with the optimization goal of:
\begin{equation}
    \begin{gathered}
        \label{equ:greedy}
        min\ depth(node) \times weight(node) \\
        weight(node) = \left\{
        \begin{array}{lcr}
        1 & & concatenation \\
        0 & & reduction \\
        \end{array} \right.
    \end{gathered}
\end{equation}
In Eq. \ref{equ:greedy}, the term depth(node) denotes the depth of a node within the tree structure. 

\subsubsection{Fault Tolerance in Wafer-Scale CIM}
\label{section:Fault}
Given the large-scale integration of Ouroboros, fault-tolerant schemes are essential to maintaining yield and reliability. Unlike selecting the "Known Good Die" in multi-chiplet architectures, a monolithic WSC requires an adaptive fault recovery mechanism that dynamically reconfigures the system to bypass failed components while preserving computational efficiency.

Ouroboros leverages a fully homogeneous compute fabric, where any core can substitute for others in its vicinity, forming a redundant replacement chain. This design allows for seamless fault recovery without disrupting inference execution. The system follows an active redundancy strategy, where all functional cores are initially active, excluding those identified as defective at fabrication. If a failure occurs during operation, two primary scenarios are considered.

\begin{figure}[htbp]
    \begin{center}
    \includegraphics[width=0.40\textwidth]{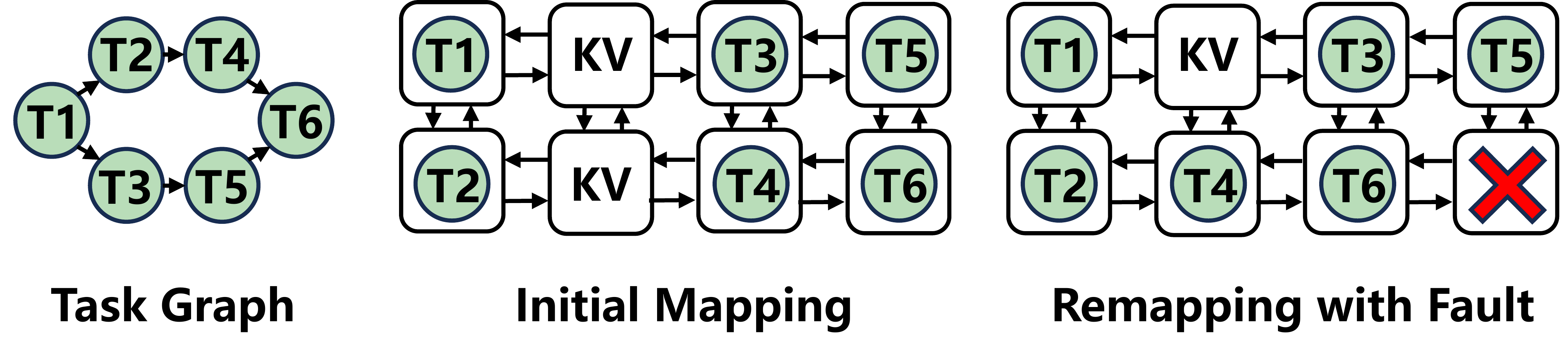}
    \end{center}
    \caption{Fault Tolerance Scheme Based on Remapping.}
    \label{fig:fault}
\end{figure}

If the failed core is responsible for KV cache storage, only the sequences stored in that core need to be recomputed, avoiding system-wide disruptions. However, if a core storing LLM weights fails, the system triggers remapping to redistribute the affected weights to adjacent cores. 
As shown in Fig. \ref{fig:fault}, we configure all cores spanning from a faulty core to the nearest core storing KV caches into a replacement chain. During this process, data in the KV cache is evicted, while weights of each core are deterministically propagated to neighboring cores and ultimately to dedicated KV cores. This recovery mechanism operates locally without invoking MIQP, completes within sub-millisecond latency, and guarantees a legal fallback.
This remapping ensures minimal impact on inference performance and prevents bottlenecks caused by memory access imbalances.

Failures are further categorized into computational unit failures and interconnect failures. In the case of a computational unit failure, the system dynamically reroutes workload execution, bypassing the defective core and redistributing the tasks across its neighbors. For interconnect failures, the routing tables within the network-on-wafer are reconfigured in real time to circumvent faulty links, ensuring continued low-latency communication. Both methodologies strictly adhere to the deadlock avoidance rule.

By integrating adaptive workload remapping and real-time routing adjustments, Ouroboros maintains high yield, robust execution, and efficient resource utilization, making wafer-scale CIM a practical solution for LLM inference.
Compared to spare-core strategies ~\cite{sun2025aphelios}, our approach maintains peak throughput at the cost of mild per-wafer variance, which can be mitigated via workload-aware placement.

\subsection{Distributed Dynamic KV Cache Management}
\label{section:DKV}
KV cache's rapid growth with context length creates a critical memory bottleneck in LLM deployments~\cite{298683,liu2024scissorhands}. Runtime-generated KV pairs' dynamic scaling also causes significant wastage under static allocation~\cite{kwon2023efficient}.
In CIM systems characterized by the tight coupling of computing and storage resources, this implies that the corresponding computing resources are also subject to wastage.
While there are existing works that have implemented dynamic management of KV cache memory on GPUs~\cite{kwon2023efficient, 298683}, they are unsuitable for our system due to the following fundamental differences:

    (1) Contrary to HBM-based KV cache solutions, Ouroboros employs a fully distributed storage system composed of small SRAM blocks interconnected by a multi-level NoC. For WSCs, the propagation of control signals in such a distributed storage system incurs high latency, making it challenging to adapt to existing centralized management methods.
    
    (2) Unlike prior work, KV cache is no longer a pure storage component. In Ouroboros, KV cache also performs in situ computations for attention. Given that CIM cannot perform computation concurrently with write operations, it is imperative to ensure load balancing during management to achieve effective separation of storage and computation. 

Therefore, it is crucial to design a distributed dynamic KV cache management method and a corresponding hardware architecture tailored for Ouroboros.

\subsubsection{Crossbar Implementation for Dynamic Management}
\label{section:CIM_Implementation}

\begin{figure}[htbp]
    \begin{center}
    \includegraphics[width=0.47\textwidth]{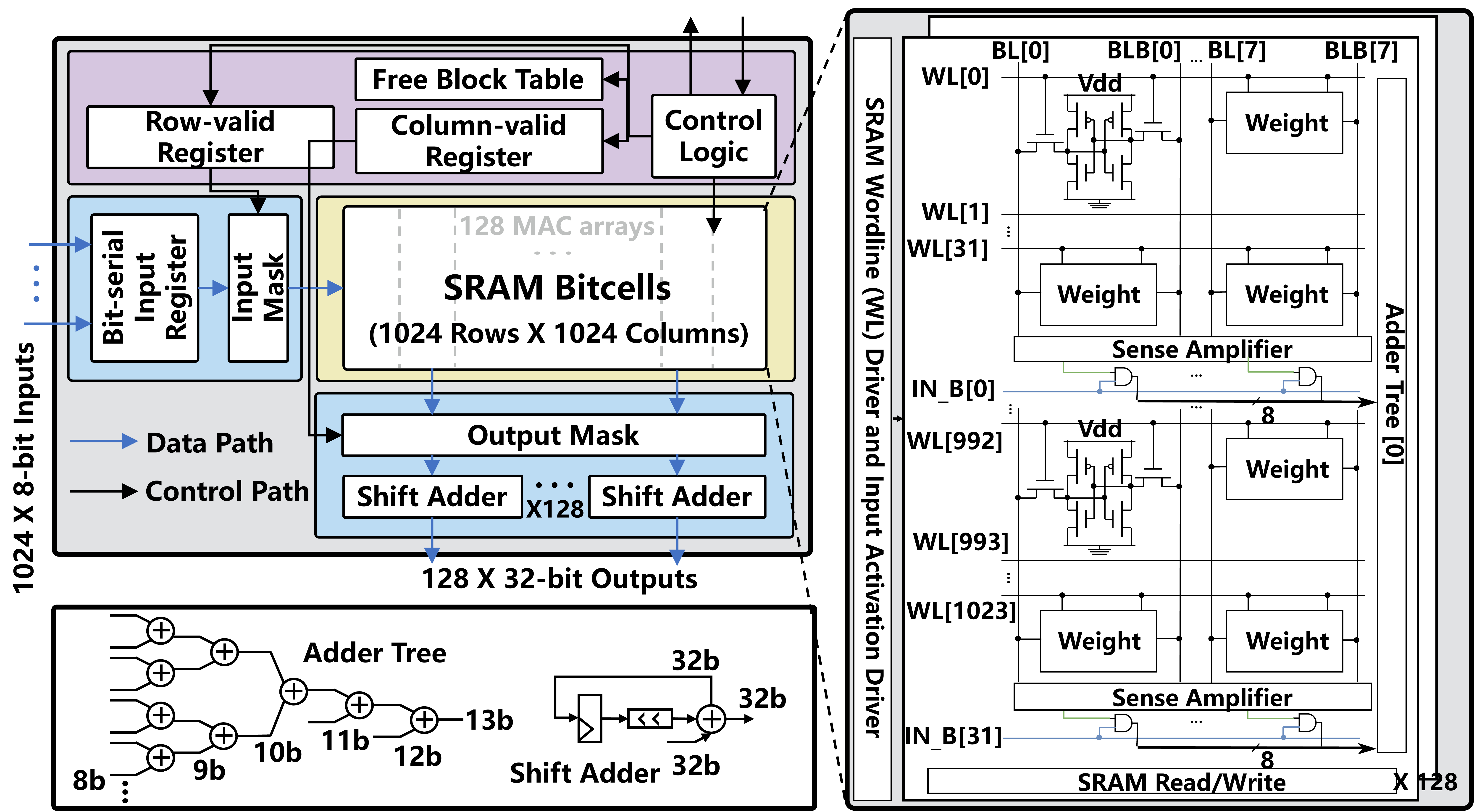}
    \end{center}
    \caption{Microarchitecture of Crossbar.}
    \label{fig:PIM_Macro}
\end{figure}

Given the high precision requirements of LLMs, we choose the digital CIM that incurs no precision loss~\cite{tu2022sdp}. The structure of a single crossbar, as depicted in Fig. \ref{fig:PIM_Macro}, comprises a $1024 \times 1024$ SRAM array, where each cell stores 1 bit, 128 shift adders, an input mask logic, an output mask logic, and a controller. The SRAM array stores $1024 \times 128$ 8-bit weights, accepts 1024 8-bit input activation, and outputs 128 32-bit partial sums.

\begin{figure}[htbp]
    \begin{center}
    \includegraphics[width=0.38\textwidth]{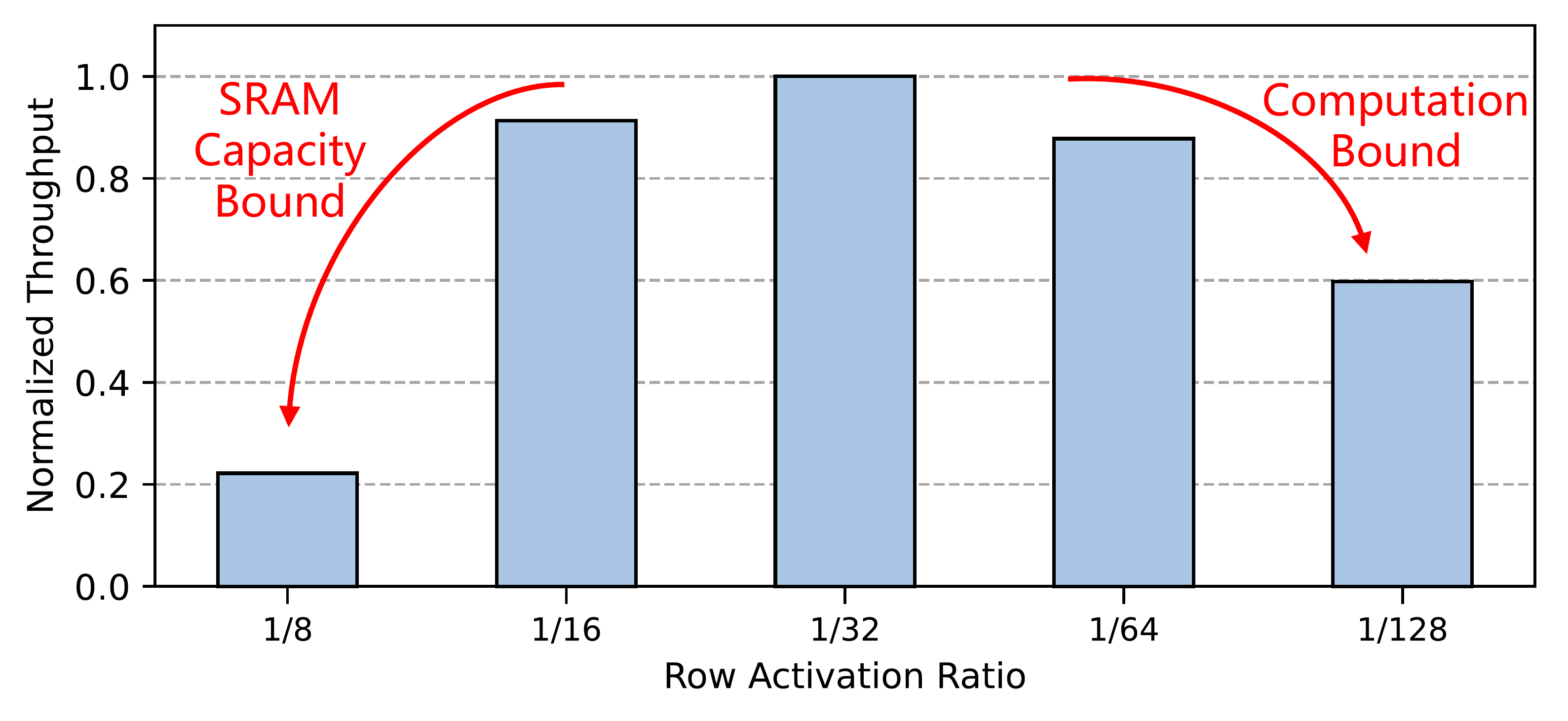}
    \end{center}
    \caption{Throughput Under Different Row Active Ratio.}
    \label{data:row_active}
\end{figure}

The SRAM CIM array includes 128 MAC arrays with a total of 1024 rows $\times$ 1024 columns of 6T bitcells for weight storage. 
Area efficiency being a primary design constraint in Ouroboros, an all-SRAM system, necessitates the use of the compact 6T bitcell rather than larger 8T or 12T bitcells designed for concurrent read and write operations.
Consequently, to maximize capacity while precluding simultaneous computation and writing, we implement the mapping and scheduling strategies detailed in Section \ref{section:kvmap}.
Wordline driver and input activation driver are shared by 128 MAC arrays. An 8:1 multiplexer selects one bit per cycle from each input activation for bitwise multiplication. MAC arrays are partitioned into 32 banks, each containing 32 rows. Each bank activates one row at a time, and 32 banks are activated simultaneously. The 1/32 sparse row activation ratio was selected to maximize SRAM area utilization while maintaining computational throughput. Fig. \ref{data:row_active} quantifies on LLaMA-13B: Higher activation ratios reduce KV cache capacity, throttling parallelism, while lower ratios deliver suboptimal computational throughput. The read data passes through the sense amplifier and is fed, along with the activation, into the bitwise multiplier (AND). The resulting partial sums are further accumulated in the adder tree. The five-stage adder tree receives 32 inputs of 8 bits each and accumulates them into a 13-bit output. This output is sign-extended to 32 bits and accumulated in the shift adder for no accuracy loss.

Each crossbar is configured to operate in either FFN mode or attention mode. In the FFN mode, the crossbar array only requires persistent storage of static weights. 
When configured in attention mode, the SRAM array is partitioned into $128 \times 1024$ logical blocks. This configuration aligns with the head dimensions of prevalent models~\cite{touvron2023llama,brown2020language,chowdhery2023palm}, avoiding costly high-precision reductions. These logical blocks are dynamically allocated to sequences based on the length of the sentences and their subsequent growth requirements. Therefore, cells are not always in an allocated state or storing valid data, necessitating the selective activation of certain rows and columns during computation. The controller performs row and column selection by controlling the corresponding masking logic through row and column-valid registers, thereby enabling a flexible dynamic allocation mechanism for the KV cache.

\subsubsection{Distributed Control}
Given that the attention mechanism of each transformer block operates independently, we manage the KV cache of each block separately.
After the inter-layer mapping has allocated the model weights,  the remaining cores are split equally between computing $Q \times K^T=S$ and $S \times V$.

\begin{figure}[htbp]
    \begin{center}
    \includegraphics[width=0.38\textwidth]{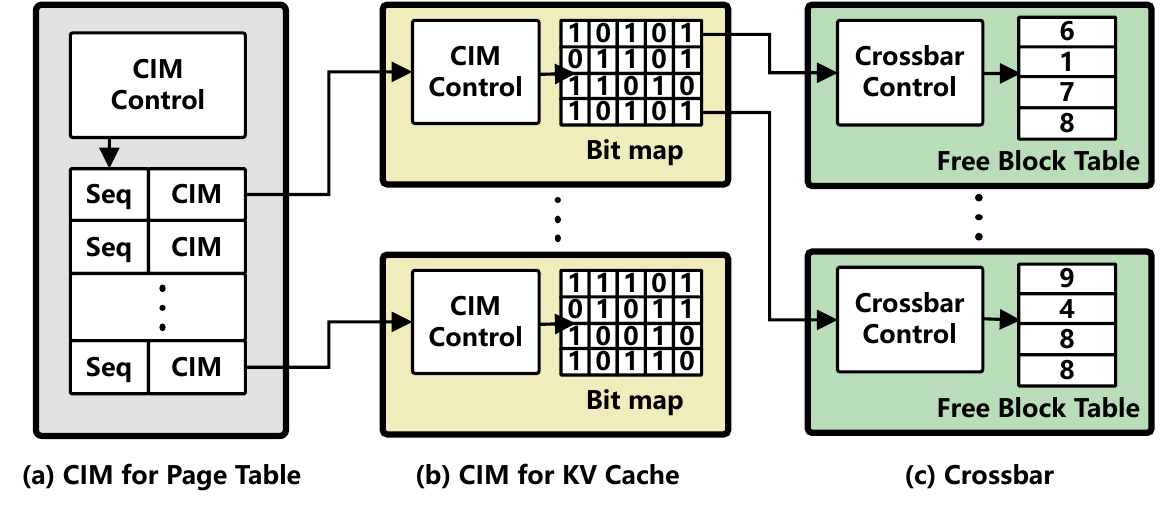}
    \end{center}
    \caption{Multi-Level Address Translation in KV Cache.}
    \label{fig:kv_map}
\end{figure}

The transformation from sequence to the physical storage location of each token is composed through a three-level mapping. As shown in Fig. \ref{fig:kv_map} (a), the first level of transformation is the page table located at the amortized storage core. As will be discussed in Section \ref{section:kvmap}, the data of each head will only be allocated to one core. Therefore, the function of the first-level page table is to transform the sequence number into a sequence of core coordinates that store each head. As shown in Fig. \ref{fig:kv_map} (b), once the core coordinates are obtained, the second level of transformation is the bitmap located in the core controller~\cite{chan1998bitmap}. The size of the bitmap is $256 \times 256$, 
entry $(m,n)=1$ denotes the $m^{th}$ sequence occupies the $n^{th}$ block, and vice versa for 0.
As shown in Fig. \ref{fig:kv_map} (c), the third level of transformation is located in the crossbar controller. Since each SRAM array contains 8 logical blocks, 8 registers are employed here to record the number of rows/columns that have been used in each logical block. These registers enable the determination of valid row and column indices. Thus, a group of cores can independently manage the KV cache without the need for centralized control.

\begin{figure*}[htbp]
    \begin{center}
    \includegraphics[width=0.85\textwidth]{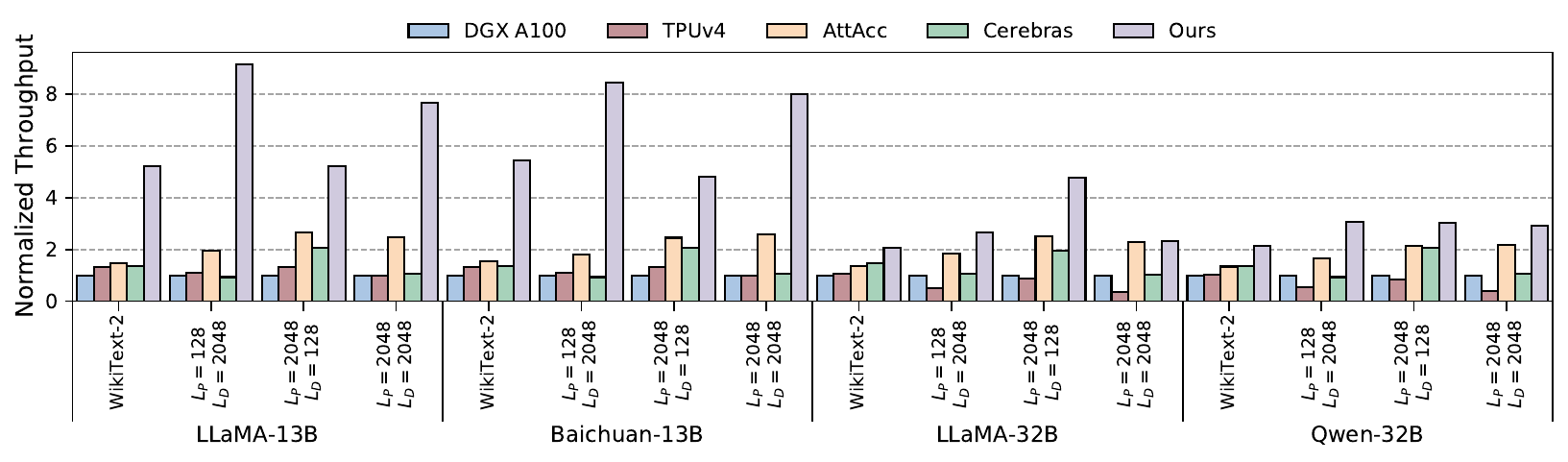}
    \end{center}
    \caption{Normalized Speedup Compared to Baselines.}
    \label{data:TC}
\end{figure*}

\subsubsection{KV Mapping}
\label{section:kvmap}
In Ouroboros, the KV cache is not merely used for storage but also needs to perform in situ in-memory computation. Therefore, meticulous design of the KV mapping is essential.

To ensure uninterrupted pipeline operation, the KV cache performs write operations for the subsequent token concurrently with the attention computation of the current token. Therefore, to ensure the separation of computation and storage, we store consecutively scheduled sequences across distinct cores. Moreover, due to the inherent independence of multiple heads in attention computations, they generate numerous independent outputs. To mitigate excessive H-tree pressure resulting from intra-core output concatenation, we distribute distinct heads across separate cores.  We number all cores used as KV cache to form a ring, recording the number of the last core allocated to the previous sequence, initially set to zero. When processing a new sequence, we start from this number and allocate a number of cores equal to the number of heads, recording the new number. When the new number exceeds the maximum core count, we start counting from zero. 

When an allocated sentence requires additional logical blocks due to length expansion. For K, the system preferentially searches for available blocks in other crossbars, as K grows along the output channel dimension and thus cannot complete accumulation within a single crossbar. Conversely, for V, the system prioritizes the allocation of available blocks within the current crossbar, as V grows along the input channel dimension, allowing for single-pass accumulation.

\subsubsection{Inter-Sequence Scheduling}
For all new inputs generated by requests, we adhere to a First-Come-First-Serve scheduling policy to ensure that no requests are starved. Preemptive scheduling can be applied to new inputs generated by auto-regression from prior requests, allowing scheduling immediately after the current input sequence is completed. Each newly scheduled request is stored in the KV cache. When the KV cache is full, the system evicts the most recently scheduled request and suspends new request scheduling until a prior request is complete. The evicted request will be placed at the front of the waiting queue.

When a new sentence enters the KV cache, which nearly fills it up, the growth of KV during the decoding phase can quickly cause the cache to become full and result in eviction, leading to computational waste. To prevent this KV cache thrashing, we set a threshold for the KV cache. 
When the remaining space in the currently indexed core drops below this threshold, the core is marked full to reserve residual capacity for KV expansion in the decoding phase.

\begin{figure*}[htbp]
    \begin{center}
    \includegraphics[width=0.85\textwidth]{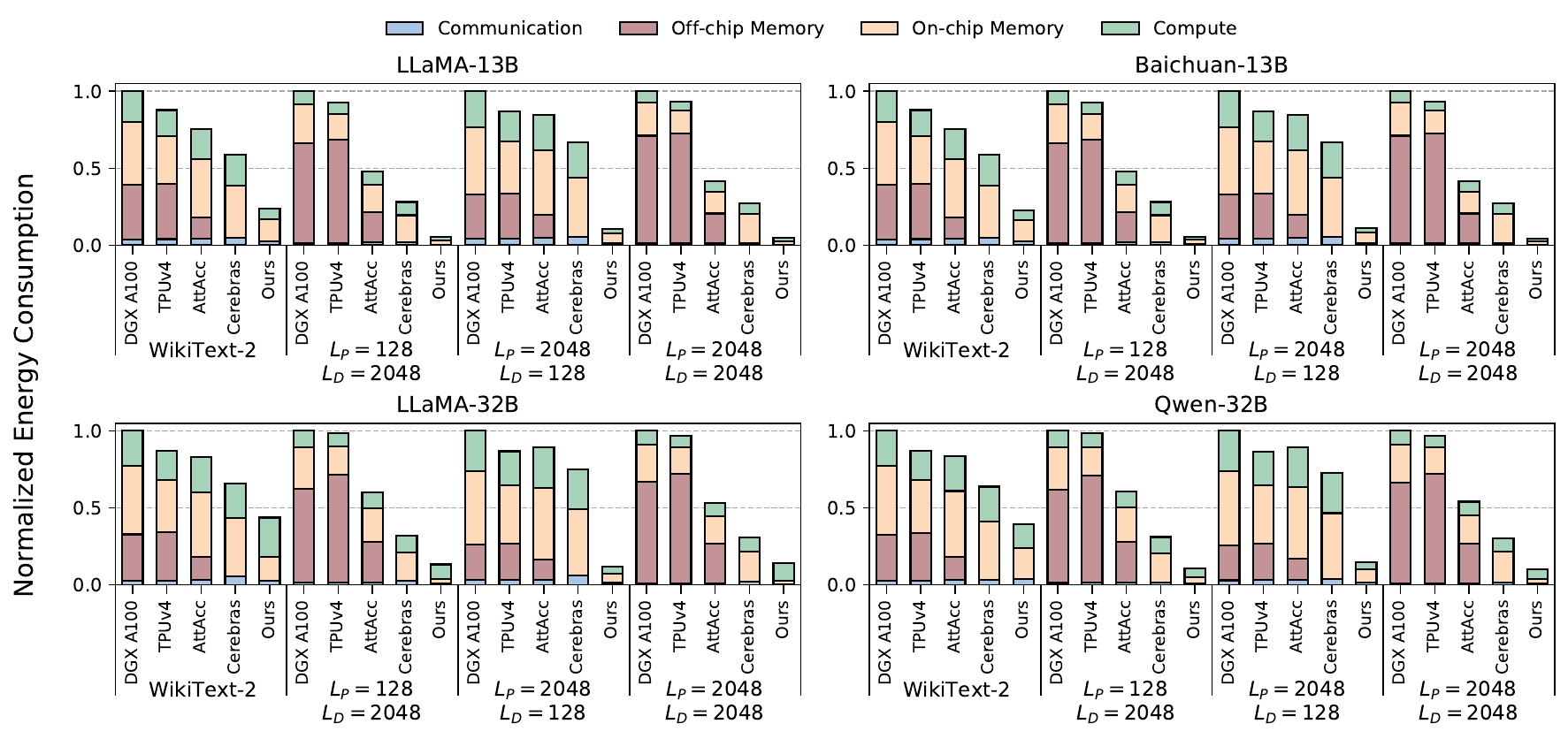}
    \end{center}
    \caption{Normalized Energy per Output Token Compared to Baselines.}
    \label{data:EC}
\end{figure*}

\section{Implementation}

We designed an E2E simulator for Ouroboros to measure performance and power consumption. The energy, area, and latency of each component are calculated as follows.

\textbf{SRAM CIM Array.} 
Our SRAM CIM array characterization using CACTI~\cite{balasubramonian2017cacti} yields an array area of 0.063$mm^2$ with average dynamic power consumption of 6.6mW and static power consumption of 0.11mW at 0.7V. Complementary logic components synthesized with Synopsys Design Compiler (DC) based on 7-nm predictive process design kit ASAP7~\cite{clark2016asap7} at 300MHz exhibit the following characteristics: AND logic gates occupy 0.0023$
mm^2$, 5-stage 32-input 8b adder trees require 0.0093$mm^2$, and 32b shift adders utilize 0.0022$mm^2$. Corresponding power measurements at RTL-level simulation (assuming 50\% input sparsity) are 0.054mW, 4.94mW, and 3.26mW respectively.
These parameters configure the validated open-source simulator MNSIM 2.0~\cite{zhu2023mnsim} with architectural specifications detailed in Section \ref{section:Architecture_Overview}, enabling derivation of core-level circuit characteristics.

\textbf{SFU \& Control.} 
We implemented the SFU and the control logic for the CIM core using Verilog, synthesized with DC based on ASAP7, operating at a frequency of 1GHz. 

\textbf{NoC.} 
We used BookSim2~\cite{jiang2013detailed}, with each router configured with eight virtual channels of depth four~\cite{zhu2024theseus}. Our NoC power characterization leverages BookSim2's baseline 32-nm models derived from ITRS2007 projections~\cite{ITRS2007}, with subsequent scaling to 7-nm technology following established methodologies~\cite{stillmaker2017scaling}. 
The inter-core bandwidth is provisioned at 256-bit bidirectional.
For inter-wafer I/O, we adopted eight 100 gigabits-per-second optical Ethernet ports~\cite{zhu2024theseus}. 

\textbf{Yield Modeling.} 
For yield $Y_r = \left(\frac{1-e^{-AD_0}}{AD_0}\right)^2 $, we modeled it using the Murphy Model~\cite{murphy1964cost} with a defect density $D_0 = 0.09/cm^2$~\cite{anandtechBetterYield} and a core area $A=2.97mm^2$. The defect core locations are randomly generated.

All aforementioned metrics are incorporated into our E2E simulation framework to execute full-system simulations.

\section{Evaluation}

\subsection{Experimental Setup}
\textbf{Workloads.}
For decoder-only models, we conducted experiments on LLaMA-13B/32B/65B~\cite{touvron2023llama}, Baichuan-13B~\cite{yang2023baichuan}, and Qwen-32B~\cite{yang2024qwen2}, which are suitable for TGP. For models that include encoder, we used T5-11B~\cite{raffel2020exploring} and BERT-large~\cite{devlin2018bert}. The dataset we utilized is WikiText-2~\cite{merity2016pointer}, a large-scale language modeling dataset derived from Wikipedia articles.

\textbf{Baseline.}
To evaluate the performance and energy efficiency of Ouroboros, we conducted a comparative analysis against SOTA hardware architectures and inference engines commonly utilized for LLM inference: 
(1) GPU Baseline: We employed a single NVIDIA DGX A100 node, equipped with eight 40GB A100 GPUs interconnected via NVLink~\cite{nvidia_a100}, a widely adopted configuration for LLM inference. For the inference engine, we employ vLLM v0.7.3~\cite{kwon2023efficient}, which enables critical optimizations including FlashAttention~\cite{dao2022flashattention} and chunked-prefill~\cite{agrawal2023sarathi}. 
(2) NPU Baseline: We utilized a cluster of eight TPU v4~\cite{jouppi2023tpu}, each with 32GB of HBM memory. For evaluations, we modified the ONNXim~\cite{ham2024onnxim} and NPUsim~\cite{kim_modsim2021} simulators to model an 8x TPUv4 system, assessing performance and power consumption separately.
(3) CIM Baseline: We adopted the DGX+AttAcc configuration as detailed in ~\cite{park2024attacc}, which includes a total of 320GB of HBM memory. 
(4) WSC Baseline: We compared Ouroboros against the Cerebras WSE-2, the most advanced wafer-scale engine, which integrates 40GB of on-chip SRAM similar to Ouroboros. 
For the inference engine, we utilize WaferLLM~\cite{he2025waferllm} and develop a cycle-accurate simulator. Validated on LLaMa-13B and 2K/4K sequences, the simulator demonstrates 6–9\% latency and approximately 10\% energy deviation.

\subsection{Performance}
Fig. \ref{data:TC} illustrates the throughput enhancement of Ouroboros compared to baseline methods when processing 1000 requests under various models and sequence length distributions, where $L_P$ denotes the prefill length and $L_D$ denotes the decode length. In comparisons involving models of size 13B, Ouroboros achieves an average throughput improvement by $5.4\times$. 
Critically, our approach exhibits scaling performance advantages with longer decode lengths, as sufficient KV cache capacity renders the memory-bound decoding phase exceptionally suited to CIM, while its intrinsic GEMV form aligns optimally with TGP.

For models of size 32B, Ouroboros achieves an average throughput improvement by $2.8\times$. 
The relatively modest enhancement relative to 13B models stems from single-wafer KV cache capacity limitations, which reduce the number of stored sentences below pipeline depth, causing pipeline underutilization.
However, this limitation can be mitigated through scaling, as detailed in Section \ref{section:Scalability}.

\begin{figure*}[htbp]
    \begin{center}
    \includegraphics[width=0.97\textwidth]{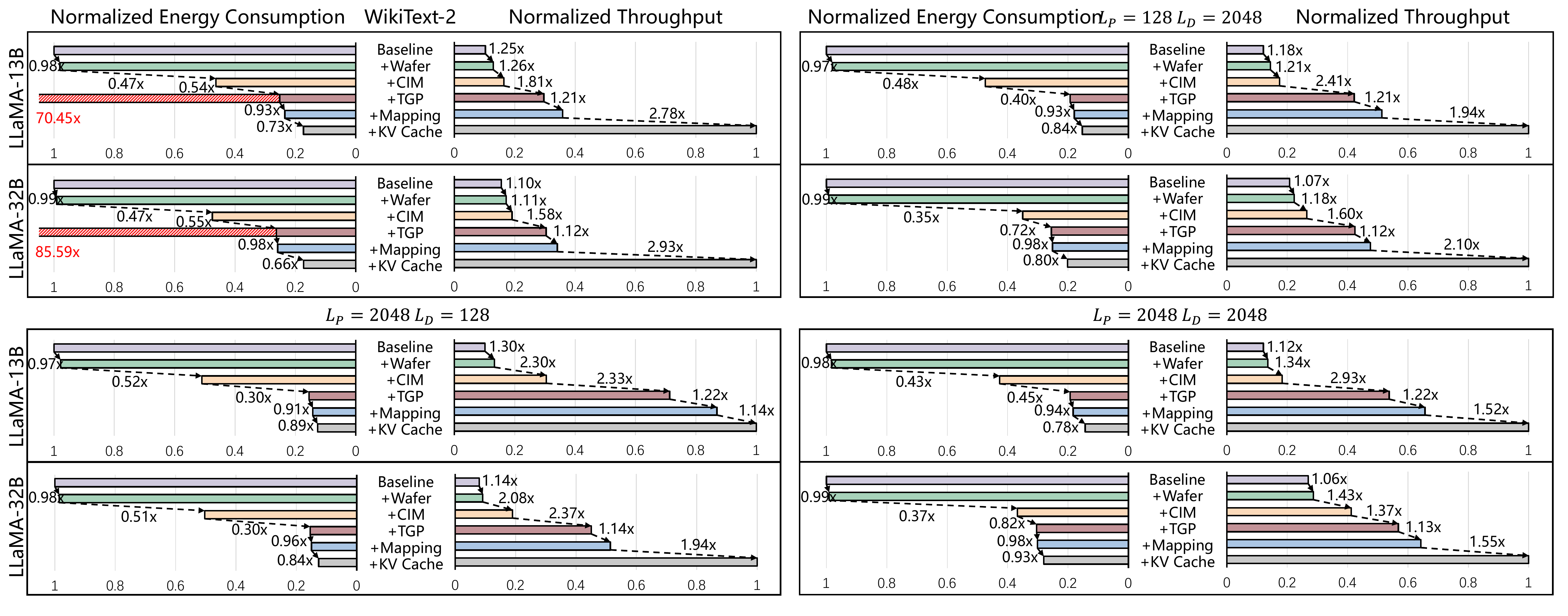}
    \end{center}
    \caption{Ablation Study on Wafer, CIM, TGP, Mapping and KV Cache Management.}
    \label{data:ablation}
\end{figure*}

\subsection{Energy Efficiency}
As shown in Fig. \ref{data:EC}, Ouroboros demonstrates superior energy efficiency, reducing energy consumption by 84\% compared to DGX A100, 82\% compared to TPUv4, 78\% compared to AttAcc, and 66\% compared to WSE2. This enhanced efficiency is attributed to two primary factors. First, Ouroboros enables the entire model to be stored on-chip, eliminating expensive DRAM accesses. Second, the adoption of full CIM for inference removes the energy overhead associated with reading weights from SRAM to the computing units. 
Ouroboros still incurs SRAM access energy from writes to I/O buffers and the KV cache.

\subsection{Adaptation of Encoder}

\begin{figure}[htbp]
    \begin{center}
    \includegraphics[width=0.42\textwidth]{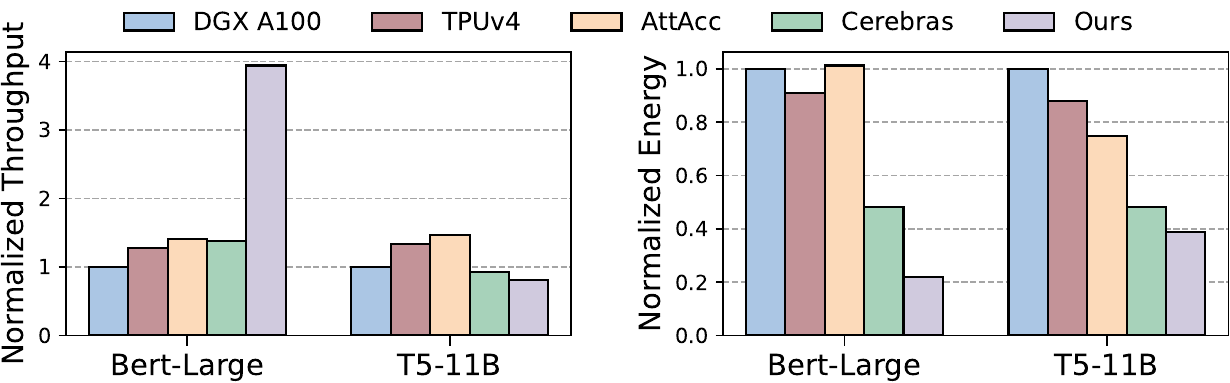}
    \end{center}
    \caption{Normalized Speedup and Energy per Output Token on Encoder-Based Model. }
    \label{data:encoder}
\end{figure}

As shown in Fig. \ref{data:encoder}, Benchmarking encoder-based models, Ouroboros achieves 3.1× and 0.7× average throughput gains over baselines for BERT-Large and T5-11B respectively.
Its less optimized performance on encoder architectures occurs because encoders only execute compute-intensive GEMM operations where memory latency is hidden in baselines, and hardware like tensor cores optimizes GEMM while CIM fundamentally operates via GEMV.
Nevertheless, TGP with blocking still significantly enhances performance, yielding an average throughput 25 times greater than that achieved with sequence-level granularity. Furthermore, our validation on four decoder-only architectures reveals only a 5\% performance reduction compared to the non-blocking version. In terms of energy consumption, 
Ouroboros continues to demonstrate substantial benefits, achieving an average reduction of 59\%.

\subsection{Ablation Study}

To evaluate Ouroboros components, we compare normalized throughput and energy consumption across configurations.
Our selected baseline comprises 64 dies interconnected in a mesh topology via NVLink ~\cite{nvidia_nvlink_2014, nvidia_nvlink_2024}, employing a model mapping scheme with tensor parallelism 8 and pipeline parallelism 8 ~\cite{liakopoulos2025iserve, liu2023janus}, alongside static KV cache management. As shown in Fig. \ref{data:ablation}, wafer-scale integration yields average throughput and energy consumption improvements of $1.15\times$ and $0.98\times$, respectively. This gain stems from field stitching's superior latency and interconnect power efficiency compared to inter-die technologies such as NVLink. When integrating CIM technology, these metrics improve to $1.49\times$ and $0.45\times$ on average, attributable to CIM's enhanced TOPS/W through in-SRAM computation.
Further incorporation of TGP elevates performance to $2.05\times$ throughput at $0.51\times$ energy consumption. TGP eliminates pipeline bubbles caused by variable sequence lengths and interleaved prefill/decoding phases – a critical optimization for deep pipelines. When evaluating TGP without CIM on WikiText (red hatched bars), average energy consumption degrades to $78.02\times$ baseline due to TGP converting all operations to GEMV, eliminating weight reuse and requiring repeated SRAM accesses per token.
Our spatial mapping strategy achieves $1.17\times$ throughput and $0.95\times$ energy consumption by minimizing high-latency multi-hop transmissions and reducing total data movement. Distributed dynamic KV cache management further improves performance to $1.99\times$ throughput and $0.81\times$ energy consumption through reduced fragmentation and increased concurrent sequence execution. This optimization proves particularly crucial for the 32B model, where a smaller KV cache capacity amplifies its benefits. Under WikiText-2 and $L_P=128$ and $L_D=2048$ settings, gains are more pronounced due to enhanced execution parallelism across mixed-length sequences and short-prefill workloads.

\subsection{KV Cache}
\begin{figure}[htbp]
    \begin{center}
    \includegraphics[width=0.38\textwidth]{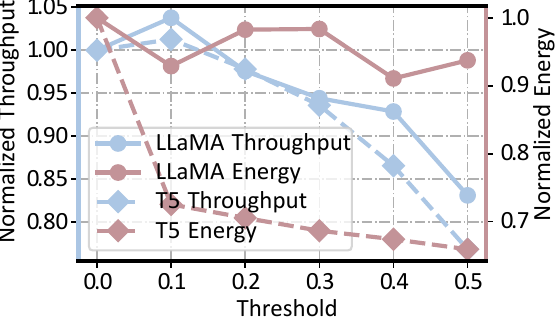}
    \end{center}
    \caption{Throughput and Energy Under Different Thresholds.}
    \label{data:threshold}
\end{figure}

Fig. \ref{data:threshold} shows the changes in Ouroboros's throughput and energy consumption under different thresholds. 
All data have been normalized based on the values at a threshold of zero. 
The red line indicates that LLaMA's energy consumption slightly decreases with fluctuations as the threshold increases, while T5's energy consumption consistently declines. This is caused by KV cache thrashing at very low thresholds, which increases recomputation and write operations. T5's more pronounced energy reduction stems from its larger attention head size, making sequences more susceptible to thrashing. Meanwhile, the blue line shows both models' throughput initially rising and then falling with increasing threshold. Reduced thrashing initially lowers recomputation and boosts throughput, but excessive thresholds decrease KV cache utilization, ultimately reducing throughput.

\subsection{Mapping}
\begin{figure}[htbp]
    \begin{center}
    \includegraphics[width=0.38\textwidth]{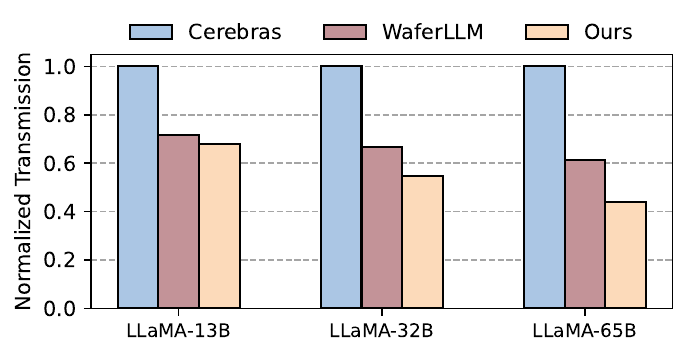}
    \end{center}
    \caption{Normalized Transmission Volume of Models with Different Scales.}
    \label{data:mapping}
\end{figure}

Our mapping algorithm, when executed on Intel(R) Xeon(R) Gold 6148 CPU @ 2.40GHz, achieves convergence within several hours. For offline mapping, this duration is considered acceptable~\cite{DBLP:conf/osdi/ZhengLZZCHWXZXG22}.
Moreover, mapping individual transformer blocks achieves a scalable design with complexity invariant to WSC count.
Fig. \ref{data:mapping} presents a comparative analysis of our proposed mapping method against other WSC mapping approaches. 
The default configuration of Cerebras employs the SUMMA algorithm~\cite{van1997summa} for GEMM computations and a pipeline all-reduce strategy~\cite{pope2023efficiently} for GEMV. Compared to Cerebras, Ouroboros reduces communication volume by 45\% on average, and by 18\% compared to WaferLLM. Furthermore, our method's advantage improves with increasing model size, as larger layers necessitate larger core arrays, exacerbating conflicts between intra-layer reductions and inter-layer communication.

\subsection{Scalability}
\begin{figure}[htbp]
    \begin{center}
    \includegraphics[width=0.38\textwidth]{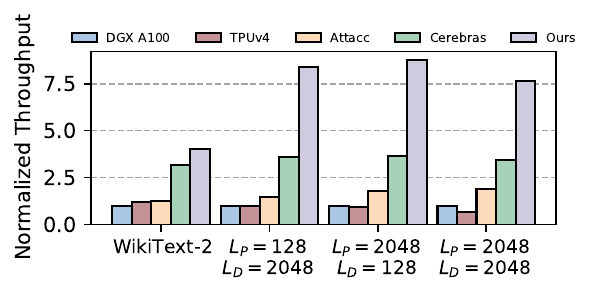}
    \end{center}
    \caption{Normalized Speedup Compared to Baselines under Multi-Wafer Scaling.}
    \label{data:TCM}
\end{figure}

\begin{figure}[htbp]
    \begin{center}
    \includegraphics[width=0.46\textwidth]{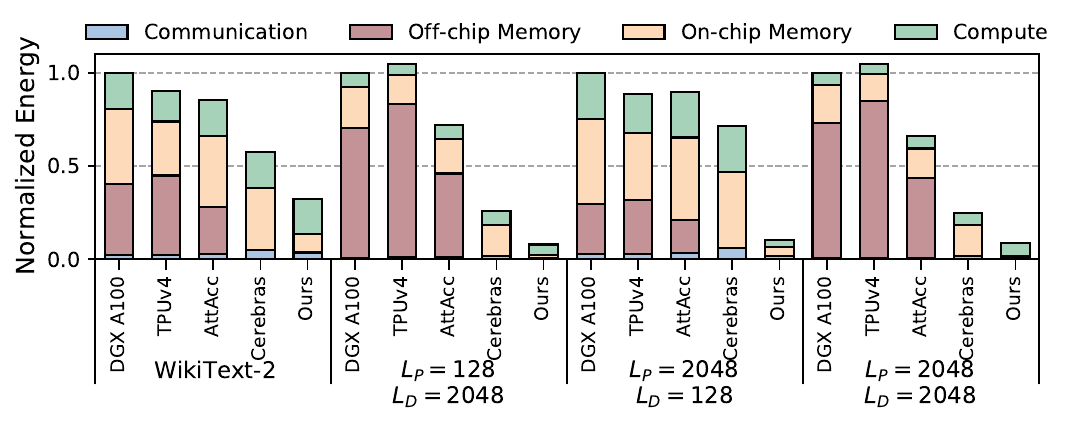}
    \end{center}
    \caption{Normalized Energy per Output Token Compared to Baselines under Multi-Wafer Scaling.}
    \label{data:ECM}
\end{figure}

To validate the scalability of our approach, we interconnected two WSCs to perform inference on the LLaMA-65B model. Fig. \ref{data:TCM} shows that the throughput of Ouroboros is, on average, $5.4\times$ that of the baseline. Additionally, Fig. \ref{data:ECM} illustrates an average energy consumption reduction of 79\%. These results underscore the excellent scalability of Ouroboros, which is a logical outcome given that larger models benefit more significantly from reduced memory access and inter-core communication overheads. Furthermore, due to the pipelined architecture, the volume of data transferred between wafers is minimal, rendering its impact on both performance and energy consumption negligible.
\label{section:Scalability}

\begin{figure*}[htbp]
    \begin{center}
    \includegraphics[width=0.90\textwidth]{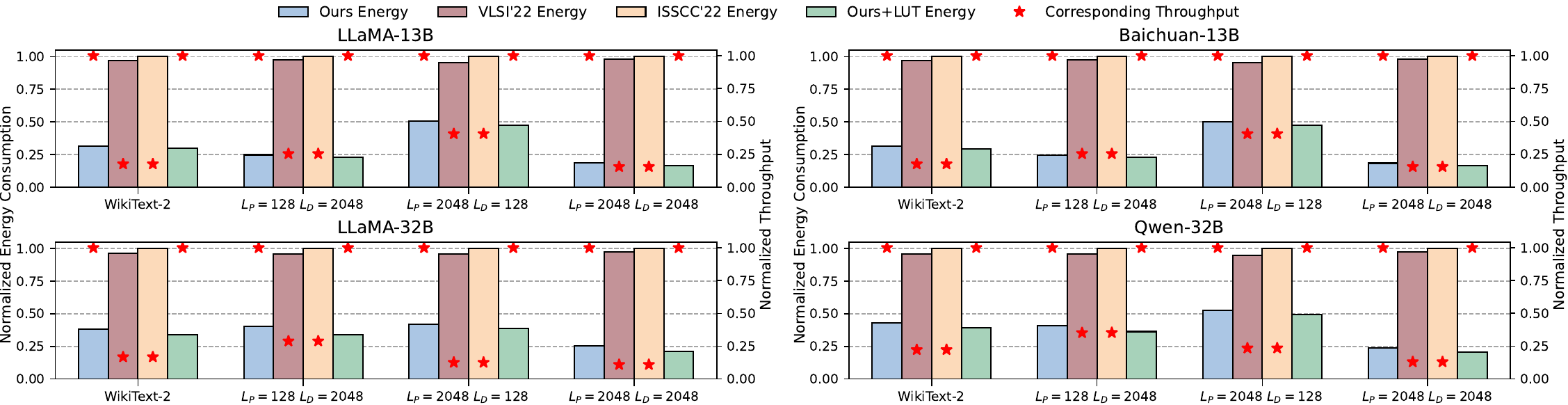}
    \end{center}
    \caption{Impacts of Different Cim Core Implementations: Applying Them to the System of Ouroboros.}
    \label{data:Cim_Core_Sys_Comparison}
\end{figure*}

\subsection{CIM Core Evaluation}
The principal contribution of Ouroboros resides in the software/hardware architecture co-design for LLM inference and its system-level optimization, rather than in pushing the circuit-level performance of CIM MAC arrays. This section positions our design relative to SOTA circuit-level approaches and explains how these techniques can be complementary in the broader hardware acceleration landscape. Shown in Table \ref{table:Cim_Core_Comparison}, baselines achieve superior TOPS/W and TOPS/mm² by focusing on aggressive circuit optimizations such as lookup table (LUT) based computations, dynamic voltage–frequency scaling, and high row-activation ratios (1/4) to boost computational density. These designs are highly effective for operator-level dense GEMM acceleration, but their limited on-chip SRAM capacity forces reliance on external DRAM for model storage and neglects the system-level design goal of E2E inference performance.

In contrast, Ouroboros targets a fundamentally different optimization point. Our LLM inference workload demands complete on-chip storage of all model weights and KV caches to eliminate the energy and latency costs of off-chip memory accesses. To achieve this, we compress the compute logic footprint in favor of maximizing SRAM capacity, adopting a conservative 1/32 row-activation ratio. This yields 5–20× more on-chip memory capacity than SOTA designs, enabling 54 GB of first-level SRAM on a single wafer—versus 2.63–11.32 GB for scaled versions of the VLSI’22 and ISSCC’22 designs. To ensure a fair comparison at the system level, we provision the baseline designs with HBM2 (1.6 TB/s) to match our memory capacity. Moreover, TOPS/W and TOPS/mm² metrics scaled to 7nm are benchmarked at 49.67 and 26.0 (VLSI'22) and 44.41 and 30.55 (ISSCC'22), respectively~\cite{davies2025defying}. As demonstrated in Fig. \ref{data:Cim_Core_Sys_Comparison}, Ouroboros achieves an average throughput improvement of $5.18\times$ and a 64\% reduction in energy consumption. This performance advantage arises because there is an intentional tradeoff for capacity-driven efficiency in memory-bound LLM inference in Ouroboros, while other solutions are fundamentally constrained by off-chip memory access. Importantly, our architecture is compatible with circuit-level optimizations: integrating LUT-based computing into Ouroboros’ cores yields an additional ~10\% energy savings, suggesting strong synergy between our system-level architecture and future circuit advances.

\begin{table}[htbp]
	\caption{Cim Core Circuit-Level Comparison}

	\begin{center}
	\resizebox{\linewidth}{!}{
		\begin{tabular}{ c | c | c | c }
			\hline
             & VLSI'22~\cite{lee202212nm} & ISSCC'22~\cite{fujiwara20225} & This work \\
            \hline 
            Technology & 12nm & 5nm & 7nm \\
            \hline
            Array size & 8Kb & 64Kb & 1024Kb \\
            \hline
            TOPS/W & $30.30$ & $63$ & $10.98$ \\
            \hline
            TOPS/$mm^2$ & $10.40$ & $55$ & $2.03$ \\
            \hline
            Wafer capacity & $2.63GB^*$ & $11.32GB^*$ & 54GB \\
            \hline
		\end{tabular}
		}
		\footnotesize{$^*$ Scaled to 7nm}\\
		\label{table:Cim_Core_Comparison}
	\end{center}
\end{table}

\section{Related Work}
\textbf{Wafer-Scale Chip. }Several existing works have designed WSC, such as UCLA\&UIUC's work~\cite{pal2021designing}, Tesla Dojo~\cite{talpes2022dojo}, Chiplet Cloud~\cite{peng2023chiplet}, Titan~\cite{yu2025cramming}, and WSC-LLM~\cite{xu2025wsc}, which utilize chiplet and advanced packaging techniques to integrate multiple chiplet into a system with an area comparable to a wafer. Cerebras~\cite{lie2022cerebras}, on the other hand, employs field stitching technology to create a complete system from an entire wafer. They all possess high computational performance and significant D2D bandwidth, allowing them to handle massive workloads. However, they still rely on DRAM or frequent SRAM read/write operations, resulting in substantial energy overhead. Several recent works have investigated wafer-scale interconnection techniques ~\cite{feng2024switch,yang2025pd,rashidi2025fred}, which remain orthogonal to the contributions of this work.

\textbf{Accelerator for Large Language Model. }
Many existing works have explored hardware architectures to accelerate Transformer models. Projects such as A3~\cite{ham20203}, ELSA~\cite{ham2021elsa}, Spatten~\cite{wang2021spatten}, Sanger~\cite{lu2021sanger}, and Energon~\cite{zhou2022energon} have employed techniques like quantization, pruning, and sparsification to speed up attention computations. However, these approaches do not address the acceleration of entire LLMs. On the other hand, DFX~\cite{hong2022dfx} and FlightLLM~\cite{zeng2024flightllm} have designed E2E LLM acceleration solutions, but they still rely on HBM or DDR, resulting in high memory access costs.

\textbf{Computing-in-Memory for Transformer. }
Numerous works have explored CIM architecture for accelerating LLM inference. TranCIM ~\cite{tu2022trancim} introduced a bitline-transpose CIM-based transformer accelerator to reduce off-chip memory access costs in attention computation. MulTCIM~\cite{tu2023multcim} leverages the sparsity in attention calculations to accelerate multimodal LLMs. 
However, these DRAM-dependent approaches compromise in situ computing efficacy. Additionally, prior CIM works optimize operator-level microarchitectures for attention/decoding, whereas our full-stack co-design introduces TGP and KV management specialized for WSC.

\section{Conclusions}
In this paper, we identify that the primary cause of the significant 'hardware scaling tax' in existing hardware architectures is the deep-level memory access. We present Ouroboros, a wafer-scale architecture comprised exclusively of SRAM CIM units, designed to reduce energy consumption by eliminating data movement. We propose TGP to mitigate low CIM core utilization caused by dynamic inputs and introduce a distributed dynamic KV cache management system with corresponding CIM hardware support to enhance KV storage utilization. Furthermore, we design a hierarchical mapping algorithm using MIQP and DP to reduce long-distance communication. Experimental results demonstrate that Ouroboros achieves significantly higher throughput ($4.1\times$) and energy efficiency ($4.2\times$) compared to SOTA architectures. 

\section*{Acknowledgments}
This work was supported in part by the National Natural Science Foundation of China (92473205, 62222411) and in part by the National Key Research and Development Program of China (2023YFB4404400). Corresponding authors are Mengdi Wang and Ying Wang.

\bibliographystyle{plain}
\balance
\bibliography{references}

\end{document}